\documentclass[twocolumn,useAMS,usenatbib]{mn2e}
\usepackage{natbib,amsmath,latexsym,amssymb}
\usepackage{epsfig,graphicx}
\usepackage{bm}
\usepackage{etex}
\newcommand{\rmd}{{\rm d}}
\newcommand{\rme}{{\rm e}} 

\newcommand{\sigmac}{\ensuremath{\Sigma_\mathrm{crit}}}

\newcommand{\seq}[1]{\begin{equation}#1\end{equation}}

\newcommand\nat{Nature}

\title[Lensing I]{Seeing in the dark -- I. Multi-epoch alchemy}
\date{\today}
\author[E. M. Huff et al.]{Eric~M.~Huff$^1$,
  Christopher~M.~Hirata$^2$, Rachel~Mandelbaum$^{3,4}$,
  David~Schlegel$^5$, \newauthor Uro\v s~Seljak$^{5,6,7,8}$, Robert~H.~Lupton$^{3}$\\
$^1$Department of Astronomy, University of California at Berkeley, Berkeley, CA 94720, USA\\
$^2$Department of Astronomy, Caltech M/C 350-17, Pasadena, CA 91125, USA\\
$^3$Department of Astrophysical Sciences, Princeton University, Peyton Hall, Princeton, NJ 08544, USA\\
$^4$Department of Physics, Carnegie Mellon University, Pittsburgh,
PA 15213, USA \\
$^5$Lawrence Berkeley National Laboratory, Berkeley, CA 94720, USA\\
$^6$Space Sciences Lab, Department of Physics and Department of Astronomy, University of California, Berkeley, CA 94720, USA\\
$^7$Institute of the Early Universe, Ewha Womans University, Seoul, Korea\\
$^8$Institute for Theoretical Physics, University of Zurich, Zurich, Switzerland
}

\begin{document}

\maketitle
\begin{abstract} 
  Weak lensing by large-scale structure is an invaluable cosmological
  tool given that most of the energy density of the concordance
  cosmology is invisible. Several large ground-based imaging surveys
  will attempt to measure this effect over the coming decade, but
  reliable control of the spurious lensing signal introduced by
  atmospheric turbulence and telescope optics remains a challenging
  problem.  We address this challenge with a demonstration that
  point-spread function (PSF) effects on measured galaxy shapes in
  current ground-based surveys can be corrected with existing analysis
  techniques.  In this work, we co-add existing Sloan Digital Sky
  Survey imaging on the equatorial stripe in order to build a data set
  with the statistical power to measure cosmic shear, while using a
  rounding kernel method to null out the effects of the anisotropic
  PSF. We build a galaxy catalogue from the combined imaging,
  characterise its photometric properties, and show that the spurious
  shear remaining in this catalogue after the PSF correction is
  negligible compared to the expected cosmic shear signal.  We
  identify a new source of systematic error in the shear-shear
  auto-correlations arising from selection biases related to masking.
  Finally, we discuss the circumstances in which this method is
  expected to be useful for upcoming ground-based surveys that have
  lensing as one of the science goals, and identify the systematic
  errors that can reduce its efficacy.
\end{abstract}


\section{Introduction}


Modern cosmologists can simulate the invisible implications of modern
cosmological models (e.g., those that can explain the cosmic microwave
background, including \citealt{2011ApJS..192...18K}) to what is generally agreed
to be a high level of precision (and probably accuracy,
c.f. \citealt{2010ApJ...713.1322L}). The easily observable
consequences of these models for observations of galaxies are not so
easy to calculate \citep[e.g.][]{2008ApJ...672...19R,2009ApJ...696..620C,2010arXiv1011.4964S},
involving as they do the physics of the familiar but nevertheless
stubbornly complicated baryons. Most of the precisely calculable
components of these models -- namely, the properties of the
distribution of dark matter on large scales in relatively linear
structures -- are not readily observable.


For the foreseeable future, the most direct observation of these
dark components is the measurement of the gravitational effects of
dark structures on the images of distant background galaxies.  These
measurements are made almost exclusively via statistical estimation of
the distortions in the ellipticities of background galaxies. This
takes advantage of the fact that galaxies have no preferred
orientation in a homogeneous, isotropic universe.


Lensing measurements have played a significant role in observational
astrophysics in the last two decades, over a range of scales and
physical regimes. Studies of galaxy evolution benefit from the ability
to understand the dark matter halos that host galaxies
\citep[e.g.][]{2004ApJ...606...67H,
  2005ApJ...635...73H,2006MNRAS.371L..60H,2006MNRAS.368..715M,
  2006MNRAS.372..758M,
  2009MNRAS.393..377M,2011arXiv1104.0928L}. Cosmologists have no other
way to directly map the large-scale matter distribution, which is
crucial for constraining models of dark energy and modified gravity
\citep{2007PhRvL..99n1302Z,2010Natur.464..256R}. On small scales, maps
of the matter distribution can be tied directly to tests of the cold
dark matter paradigm and simulations of the formation and evolution of
dark matter halos.

Much has been made of the scientific potential of this technique. Five
years ago, weak lensing was identified by the Dark Energy Task Force
\citep{2006astro.ph..9591A} as the most promising tool for
constraining cosmological models. Several large ground-based and
space-based survey proposals place a weak lensing measurement among
their primary science drivers, including the Panoramic Survey
Telescope and Rapid Response System (Pan-STARRS)\footnote{\tt
http://pan-starrs.ifa.hawaii.edu/public/}, the Dark Energy Survey
(DES)\footnote{\tt http://www.darkenergysurvey.org/}, the Hyper
Suprime-Cam (HSC, \citealt{2006SPIE.6269E...9M}) survey, the Large
Synoptic Survey Telescope (LSST)\footnote{\tt http://www.lsst.org/},
Euclid\footnote{\tt http://sci.esa.int/euclid/}, and the Wide-Field
Infrared Survey Telescope (WFIRST)\footnote{\tt
http://wfirst.gsfc.nasa.gov/}.

For all the promise, the technical challenges for these future
experiments remain formidable. An order-unity distortion to background
galaxy images is produced by a physical, projected matter overdensity
of
\begin{equation} 
  \sigmac=\frac{c^2}{4\pi G}\frac{d_S}{d_L d_{LS}},
\end{equation} 
where $d_L$, $d_S$, and $d_{LS}$ are the angular-diameter distance
from the observer to the lens and source, and from the lens to the
source, respectively.  For characteristic distances of approximately a
Gpc, the critical surface density is $0.1\,$g$\,\,$cm$^{-2}$.  Typical
fluctuations in the matter density field projected over cosmological
distances are a thousand times smaller than this, so order 10
Mpc-scale density fluctuations in the universe will typically produce
changes in galaxy ellipticities of order $e \approx 10^{-3}$ to
$10^{-2}$ in magnitude. In the shot-noise dominated regime, the
leading-order contribution to the variance in the correlation function
of the ellipticity distortions is
\begin{equation}
\mathrm{Var}\left(\xi_{\epsilon}\right)=\frac{\sigma_{\epsilon}^4}{N_\mathrm{pair}^2}.
\end{equation} 

For a shallow ($\langle z \rangle=0.5$) galaxy survey
with shape noise due to random galaxy ellipticities $\sigma_{\epsilon} \approx 0.3$ and 100
$\mathrm{deg}^2$ of sky coverage, reducing the shot noise contribution
below the expected cosmological signal requires a surface density of
usable source galaxies of at least a few per square arcminute.

Worse, for ground-based imaging surveys, the observed shape
distortions arising from atmospheric turbulence and optical
distortions from the telescope are typically of order several percent,
with coherence over angular scales comparable to that of the lensing
shape distortions. A competitive measurement of the amplitude of
matter fluctuations requires suppressing or modeling these coherent
spurious distortions to of order one part in $10^3$, and future surveys
will need to do a factor of several better.

Achieving both the statistical precision and control of systematic
errors that is required
for such a measurement has proved to be a challenge. The early
detections \citep{2000MNRAS.318..625B, 2000A&A...358...30V,
  2001ApJ...552L..85R, 2002ApJ...572...55H, 2003MNRAS.341..100B,
  2003AJ....125.1014J} showed the promise of the method and confirmed
the existence of lensing by large-scale structure at roughly the
expected level. However, they also highlighted some of the systematic
errors: in particular, $B$-mode shear (which cannot be produced by
lensing at linear order and is thus indicative of systematic effects)
was present at a sub-dominant but non-negligible level. Since then, the
weak lensing community has moved in the direction of both deep/narrow
surveys with the {\slshape Hubble Space Telescope} ({\slshape HST})
and wide/shallow surveys on the ground. The Cosmological Evolution
Survey (COSMOS) is the premier example of the former: in addition to
2-point statistics \citep{2007ApJS..172..239M, 2010A&A...516A..63S},
it has also produced three-dimensional maps of the matter distribution
\citep{2007ApJS..172..239M} and the lensing 3-point correlation
function \citep{2011MNRAS.410..143S}. Excellent control of lensing
systematics in COSMOS was also achieved thanks to the small number of
degrees of freedom controlling the PSF (mostly focus variation;
\citealt{2007ApJS..172..203R}) and detailed modeling of charge
transfer inefficiency \citep{2010MNRAS.401..371M}. However, COSMOS
covers only 1.6 deg$^2$, and the small field of view of {\slshape HST}
instruments makes significantly larger surveys impractical. The
principal recent ground-based cosmic shear programme has been the
Canada-France-Hawaii Telescope Legacy Survey (CFHTLS). There are now
several cosmic shear results from different subsets of the CFHTLS data
\citep{2006A&A...452...51S, 2006ApJ...647..116H, 2007MNRAS.381..702B,
  2008A&A...479....9F}, and the CFHT lensing team is completing a
reanalysis using recent advances in PSF determination and galaxy shape
measurement.

In light of the efforts shortly to be made by large, expensive surveys
to measure cosmic shear, we consider it imperative to show that such a
measurement can be performed accurately, without significant
contaminating systematic errors, from a ground-based observatory.
This goal includes doing a cosmic shear measurement with each of the
wide-angle optical surveys that presently exist.  To this end, we have
re-coadded the repeat observations on the equatorial stripe (stripe
82) of the Sloan Digital Sky Survey (SDSS), using methodology that
will optimise these new coadds for precision shear measurements. Our
goal is to reduce the systematic errors arising from uncorrected PSF
anisotropies below the statistical errors. We begin by specifying our
requirements in Sec.~\ref{sec:design}, and describing the data that we
use in Sec.~\ref{sec:data}.  A description of the coaddition and
catalogue-making pipeline follows in Section \ref{sec:algorithms}.  We
describe our method for estimating two-point functions of star and
galaxy shapes in Sec.~\ref{sec:cf_estimation}. Demonstrations of the
data quality and suitability for sensitive weak lensing measurements
are described in Section~\ref{sec:diagnostics}. We conclude with
lessons for future experiments in Sec.~\ref{sec:discussion}.

\section{Design Requirements}\label{sec:design}

Weak lensing measurements on large scales are vulnerable to a variety
of systematic measurement errors. In order to measure cosmic shear on
the scales described above, we must first have a clear idea of what
the possible sources of these systematic errors are, and to what level
(quantitatively) they must be suppressed. This section describes in
turn the most common generic sources of measurement error relevant for
weak lensing, and lays out quantitative methods for detecting their
presence in our final catalogue.

The PSF\footnote{Here we use the term ``PSF'' to denote all
contributions: the atmosphere, optics, tracking errors, charge
diffusion, and pixelization.} of the SDSS survey exhibits significant
spatial and temporal variations across the entire survey. 
We model these effects as a spatially-varying convolution kernel
$G$. The observed image $I({\bmath x})$ at some position ${\bmath x}$
is related to the ``true'' image $f$ by
\begin{equation} 
I \left({ \bmath x} \right) = \int f({\bmath y}) G({\bmath
x}-{\bmath y})\, \rm d^2{\bmath y},
\end{equation} 
where $G$ is the convolution kernel appropriate to the
region of sky under examination.

One effect of a spatially-varying PSF $G$ is to produce a spurious
shear field determined by the atmosphere and telescope that is
statistically independent of and superposed upon the undistorted
galaxy shape pattern. Point sources (stars and completely unresolved
galaxies -- we have no need, at present, to distinguish these) 
sample only the field sourced by $G$, and so can be used to
constrain a model for the systematics field. Any uncorrected additive
shear contribution due to the ellipticity of $G$ will produce a
correlation between the measured galaxy and point-source shapes. This
additive shear will be statistically uncorrelated with the true cosmic
shear signal.

The masking steps of the catalogue construction procedure can also
produce a significant shape selection bias. For the photometric
pipeline used here, masked regions are defined as sets of pixels; a
galaxy is rejected from the catalogue if the set of pixels making up a
galaxy intersects the set of masked pixels. On the masked region
boundary, galaxies aligned across the mask boundary are more likely to
be rejected from the catalogue than galaxies aligned along it producing
a spurious shear. This will affect both stars and galaxies, but the
effect on spurious galaxy ellipticities will be much larger than that
on stars (as the dispersion in measured stellar ellipticities is very
small). This mask selection bias produces an additive shear, which
will also be statistically uncorrelated with the true cosmic shear
signal.

These two effects enter together as an additive term in the shape clustering statistics, as
\begin{equation} 
  \xi_{\rm measured}(\theta) = \xi_{\rm cosmic}(\theta)+\xi_{\rm systematics}(\theta).
\end{equation} 

The point-source and galaxy populations have different sensitivities
to the ellipticity of the PSF, to optical distortions, and to the geometry of the mask. If
these are accounted for, a measurement of the point source-galaxy
cross correlation provides a straightforward estimate of the spurious
signal sourced by uncorrected PSF variation\footnote{This statement is
  true for sufficiently large areas that any chance superpositions of
  PSF ellipticity and the lensing shear average out.  For this reason,
  we impose this test on chunks with area $\ge 25$ deg$^2$.}. We will
require that the amplitude of this spurious correlation in our final
shape catalogue be sub-dominant to the statistical errors -- in
particular, that the additive PSF systematics amplitude be constrained
to less than the statistical errors.

The average ellipticity measured for the gravitationally sheared
images of a population of galaxies is proportional to the applied
shear; the exact value of this calibration depends on the surface
brightness profiles of the galaxies. We will address the shear
calibration uncertainties in a companion paper.

\section{Data}\label{sec:data}

\subsection{The Sloan Digital Sky Survey}

The Sloan Digital Sky Survey (SDSS; \citealt{2000AJ....120.1579Y}) and
its successor SDSS-II \citep{2008AJ....135..338F} mapped 10000 square
degrees across the north galactic cap using a dedicated wide-field 2.5
m telescope at Apache Point Observatory in Sunspot, New Mexico
\citep{2006AJ....131.2332G}. The SDSS camera, described in
\citet{1998AJ....116.3040G}, images the sky in five optical bands
({\it u,g,r,i,z}; \citealt{1996AJ....111.1748F, 2002AJ....123.2121S})
with the charge-coupled device (CCD) detectors reading out at the
sidereal rate. Each patch of sky passes in sequence through the five
filters (in the order {\it r,i,u,g,z}) along one of the six columns of
mosaicked CCDs, and is exposed once in each filter for 54.1 s. The
site is monitored for photometricity \citep{2001AJ....122.2129H,
  2006AN....327..821T}. Data undergo quality assessment
\citep{2004AN....325..583I}, and final calibration is done using the
``ubercalibration'' procedure based on photometry of stars in run
overlap regions \citep{2008ApJ...674.1217P}.  We use the data from the
seventh SDSS data release \citep{2009ApJS..182..543A}, with an updated
calibration from the subsequent data release.

The footprint of one night's observing is six columns of imaging the
width of one CCD (13.52 arcmin) separated by slightly less than one
CCD width (11.65 arcmin). Imaging taken during a continuous period of
time on one night is collectively termed
a {\it run}; each separate column of imaging is, sensibly, a {\it
camera column} (or ``camcol''), and the imaging along each camera
column is chopped for processing purposes into 8.98 arcmin long {\it
frames} or {\it fields}. Successive runs are interleaved, in order to
fill in the gaps between camera columns. Pairs of interleaved runs
along the same great circle are {\it stripes} (each of which has a
north and a south {\it strip}).

\subsection{Stripe 82}

Most of the SDSS imaging data were acquired in the northern galactic
cap, with galactic latitude $\left|b\right|>30$. For commissioning,
and during sidereal times when the primary survey region was
unavailable, the telescope frequently imaged a 2.5 degree wide stripe
of sky along the celestial equator with right ascension (RA) in the
interval $-50<{\rm RA}<+50^\circ$. The SDSS-II supernova project
\citep{2008AJ....135..338F} observed this region many times during the
months of September--November over the years 2005--2007
 to collect multicolour light curves of Type Ia
 supernovae. In the survey nomenclature, this region is Stripe
82. At any given location on the Stripe, there are on average 120
contributing interleaved imaging runs, comprising in aggregate almost
as much imaging data as exist in the remainder of the combined SDSS-I
and SDSS-II footprint. It is here that significant gains can be made
from image coaddition.

\subsection{Single-epoch data processing}

The raw imaging data is processed by the automated SDSS photometric
pipeline, {\sc Photo} \citep{2001ASPC..238..269L}. This pipeline has
components to handle astrometric and photometric calibration as well
as catalogue construction; it also generates an array of data quality
measurements describing the telescope point-spread function (PSF), the
locations of unreliable pixels, and measurements of the photometric
quality of individual frames. Many of these data quality indicators
are used during the construction of the coadd imaging and its
associated catalogue. Their use is described below. A detailed
description of the image processing pipeline and its outputs can be
found in \citet{2002AJ....123..485S}.  Outputs can be found in
locations specified by the SDSS data release papers
\citep{2003AJ....126.2081A, 2004AJ....128..502A, 2005AJ....129.1755A,
2006ApJS..162...38A, 2007ApJS..172..634A, 2008ApJS..175..297A,
2009ApJS..182..543A}.

{\sc Photo} produces a number of intermediate outputs for the
single-epoch SDSS imaging that we use in the coaddition process.
Corrected Frames, or {\tt fpC} files, are produced by the pipeline
from the raw CCD images of single frames; these are bias-subtracted
and flat-fielded, and a non-linearity correction is applied where
appropriate. These are the images that are combined during the
coaddition process below.

{\sc Photo} also generates a bitmask (an {\tt fpM} file) for each
frame describing pixels that are known to be defective. Pixels are
marked in this bitmask as saturated, cosmic-ray contaminated,
interpolated (if a column or pixel is known to be saturated, or is
{\it a priori} marked as unreliable, {\sc Photo} interpolates over
that region). We use these bitmasks to exclude bad pixels from the
image coaddition.

Astrometric solutions ({\tt asTrans} files) are produced by {\sc
Astrom} for each SDSS frame. Systematic errors in the astrometric
positioning are reported to less than 50 mas, and the relative
astrometry between successive overlapping frames is approximately 10
mas \citep{2003AJ....125.1559P}.  The astrometric solution for each run
\citep{2003AJ....125.1559P} is determined by matching against
astrometric standard stars from the USNO CCD Astrograph Catalogue
\citep[UCAC][]{2000AJ....120.2131Z} catalogue. 
The coaddition algorithm relies on the astrometric solutions provided;
we have found it unnecessary to re-solve the astrometry.

For photometric calibration, we use the ``ubercal'' solutions derived
by \citet{2008ApJ...674.1217P}.

The SDSS pipeline uses bright, isolated stars with apparent magnitudes
brighter than $19.5$ to construct a model of the PSF and its variation
across each frame. For each frame, the stellar images for the three
neighbouring frames along the scan in both directions are used to
produce a set of Karhunen-Lo\`eve (KL) eigenimages describing the PSF
variation \citep{2001ASPC..238..269L}. A global PSF model for the
frame is constructed by allowing the first few KL components to vary
up to second order in the image coordinates across the frame, with the
coefficients of the variation being tied to the aforementioned bright
stars. The KL eigenimages and coefficients of their spatial variation
are stored by {\sc Photo} for each band in the {\tt psField}
files. These are taken as inputs to the coaddition process and used
for PSF correction. We will test the fidelity of this PSF model in the
coadded images on stars that were too faint to perform a reliable PSF
determination in the single-epoch data.

\section{Algorithms}
\label{sec:algorithms} 

\begin{figure}
\includegraphics[width=\columnwidth]{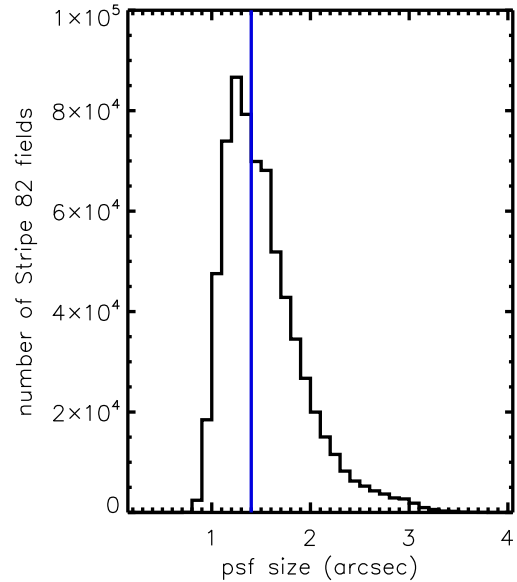}
\caption{\label{fig:psf_sizes}The distribution of PSF FWHM in the
  {\it r} band for all frames on Stripe 82. The half-width of the
  target PSF after rounding is indicated by the solid vertical line.}
\end{figure}

Our general strategy for correcting for the effects of seeing is
similar to that suggested in \citet{2002AJ....123..583B}. We will
apply a rounding kernel to each single-epoch image prior to stacking
the ensemble. The large variation in SDSS PSF sizes (see
Fig.~\ref{fig:psf_sizes}) will require a trade-off between rejection
of a large fraction of the available imaging, and significant dilution
of the signal due to the rounding convolution. Stacking the images
without a kernel, however, will produce a PSF with large variations
-- including {\em steps} at run boundaries or the edges of regions masked
due to e.g. cosmic rays -- that will be difficult to model accurately.

\subsection{Field smoothing}
\label{sec:smooth}

This section describes the operation of smoothing the map so as to
make the effective PSF equal to some target PSF.  Here we will denote
the intrinsic PSF of the telescope by $G({\bmath x})$, so that if the
intrinsic intensity of an object on the sky is $f({\bmath x})$, the
actual image observed is 
\seq{I( { \bmath x}) = \int G({\bmath
y})f({\bmath x}-{\bmath y})\rm d^2{\bmath y}\equiv [G\otimes f]({\bmath
x}).\label{eq:igf1}}
 Of course this image is only sampled at values of
${\bmath x}$ corresponding to pixel centres.  Our principal objective
here is to construct the kernel $K$ such that \seq{[I\otimes
K]({\bmath x}) = [\Gamma\otimes f]({\bmath x}) {\rm ~~or~~} [G\otimes
K]({\bmath x}) = \Gamma({\bmath x}),\label{eq:igf}} where $\Gamma$ is
the target PSF.  In order to do this, we need to first choose a target
PSF $\Gamma$ and then determine the appropriate convolution kernel
$K$, which will differ for every imaging run contributing to the
coadds at a given position depending on the full position-dependent PSF model in each
run.  These are the subjects of Secs.~\ref{ss:gamma} and \ref{ss:k}
respectively.

\subsubsection{The target PSF}
\label{ss:gamma}

Here we consider the target PSF $\Gamma$.  It must be constant across
different runs in order for the co-add procedure to make sense,
although it need not be the same in different filters.  There is a
large advantage in having $\Gamma$ be circularly symmetric.  Gaussians
are convenient since most galaxy shape measurement codes are based on
Gaussian moments, but this is not a requirement.  In fact the PSF $G$
delivered by most telescopes, including the SDSS, has ``tails'' due to
the atmosphere at
large radius that are far above what one could expect from a Gaussian.
These can in principle be removed by a convolution kernel $K$ that has
negative tails at large radius, but there are problems when these
tails extend to the field boundaries or across bad columns in the CCD.
Therefore we have chosen the double-Gaussian form for $\Gamma$:
\seq{\Gamma({\bmath x}) = \frac{1-f_{\rm
w}}{2\pi\sigma_1^2}\rme^{-{\bmath x}^2/2\sigma_1^2} + \frac{f_{\rm
w}}{2\pi \sigma_2^2}\rme^{-{\bmath x}^2/2\sigma_2^2}} with
$\sigma_2>\sigma_1$.  This functional form manifestly integrates to
unity, and has a fraction $f_{\rm w}$ of the light in the ``large''
Gaussian.  The two Gaussians have widths $\sigma_1$ and $\sigma_2$,
respectively, with $\sigma_1<\sigma_2$.

The parameters of the double-Gaussian were adjusted by trial and error
so that a compactly supported kernel $K$ ($13\times 13$ pixels) can
achieve $G\otimes K\approx\Gamma$ for a wide range of real PSFs $G$
delivered by the SDSS.  The most critical parameter is the width of
the central Gaussian, $\sigma_1$.  This is the main parameter
controlling the seeing of the final co-added image: if it is set too
high then many galaxies become unresolved, whereas if it is set too
low then a large number of fields with moderate seeing will have to be
rejected because it will be impossible to find a kernel $K$ that
achieves the target PSF without dramatically amplifying the noise.

The PSF size distribution in the {\it r} band is shown in
Fig.~\ref{fig:psf_sizes}.

\begin{table*}
\caption{\label{tab:params}Parameters for the PSF repair in different
filters.}
\begin{tabular}{lcccccccccccc} 
\hline\hline 
Parameter & & $u$ & & $g$ & & $r$ & & $i$ & & $z$ & & Units \\ 
\hline \multicolumn{13}{c}{Target PSF parameters} \\ 
\hline 
$\sigma_1$ ({\tt PSF\_SIZE}) & & 1.80 & & 1.40 & & 1.40 & & 1.40 & & 1.40 & & pixels \\ 
$\sigma_2$ ({\tt PSF\_SIZE\_WING}) & & 5.10 & & 5.10 & & 5.10 & & 5.10 & & 5.10 & & pixels \\ 
$f_{\rm w}$ ({\tt FRACWING}) & & 0.035 & & 0.035 & & 0.035 & & 0.035 & & 0.035 & & pixels \\ 
FWHM of target PSF $\Gamma$ & & 1.68 & & 1.31 & & 1.31 & & 1.31 & & 1.31 & & arcsec \\ 
50 per cent Encircled Energy Radius & & 0.86 & & 0.67 & & 0.67 & & 0.67 & & 0.67 & & arcsec \\
\hline \multicolumn{13}{c}{Kernel acceptance parameters} \\ 
\hline {\tt CUT\_L2} & & 0.001 & & 0.0025 & & 0.0025 & & 0.0025 & & 0.0025 & & \\ 
{\tt CUT\_OFFSET} & & 0.04 & & 0.01 & & 0.01 & & 0.01 & & 0.01 & & \\ 
{\tt CUT\_ELLIP} & & 0.002 & & 0.0005 & & 0.0005 & & 0.0005 & & 0.0005 & & \\ 
{\tt CUT\_SIZE} & & 0.01 & & 0.0025 & & 0.0025 & & 0.0025 & & 0.0025 & & \\ 
{\tt CUT\_PROF4} & & 0.04 & & 0.01 & & 0.01 & & 0.01 & & 0.01 & & \\ 
\hline \multicolumn{13}{c}{Co-addition parameters} \\ 
\hline {\tt DELTA\_SKY\_MAX1} & & 0.5 & & 0.25 & & 0.25 & & 0.25 & & 0.25 & & nmgy arcsec$^{-2}$ \\ 
{\tt DELTA\_SKY\_MAX2} & & 0.04 & & 0.02 & & 0.02 & & 0.02 & & 0.02 & & nmgy arcsec$^{-2}$ \\ 
\hline\hline
\end{tabular}
\end{table*}

\subsubsection{The convolution kernel and its application}
\label{ss:k}

Equation~(\ref{eq:igf}) can formally be solved in Fourier space by
taking the ratio, $\tilde K({\bmath k}) = \tilde\Gamma({\bmath
k})/\tilde G({\bmath k})$, where the tilde denotes the Fourier
transform and ${\bmath k}$ the wave vector.  Unfortunately, this
simple idea comes with two well-known problems.  One is that if the
PSF has power only up to a certain wave number $k_{\rm max}$, then it
is impossible to divide by $\tilde G({\bmath k})=0$.  The other is
that the PSF varies slowly across the field, i.e. $G$ in
Eq.~(\ref{eq:igf}) formally depends on ${\bmath x}$ as well as
${\bmath y}$.

The solution to the first problem is that instead of taking a simple
ratio in Fourier space, we minimise the $L^2$ norm of the error,
\seq{{\cal E}_1 = \int |\Gamma({\bmath x})-[G\otimes K]({\bmath x})|^2
  \,\rmd^2{\bmath x} \equiv \|\Gamma-G\otimes K\|^2,} subject to a
constraint on the $L^2$ norm of the kernel: 
\seq{{\cal E}_2 = \int
  |K({\bmath x})|^2 \rmd^2{\bmath x} \equiv \|K\|^2.}  
If the input
noise is white (which is a good approximation), then the noise
variance on an individual pixel in the convolved image is "${\cal
  E}_2$ times the noise variance in the input image.  Roughly
speaking, for kernels that attempt to ``deconvolve'' the original PSF,
and consequently have large positive and negative contributions,
${\cal E}_2$ will come out to be very large.  We adopt a requirement
that ${\cal E}_2\le 1$.  For kernels that poorly approximate the
target PSF $\Gamma$, ${\cal E}_1$ will be very large.  The problem of
minimising ${\cal E}_1$ subject to a constraint on ${\cal E}_2$ is
most easily solved by transforming to the Fourier domain and then
using the method of Lagrange multipliers: \seq{\tilde K(\bmath k) =
  \frac{\tilde G^\ast({\bmath k})\tilde\Gamma({\bmath
      k})}{|\tilde G({\bmath k})|^2+\Lambda}.
  \label{eq:kg}} Here the positive real number $\Lambda$ is the
Lagrange multiplier and its value is adjusted until ${\cal
  E}_2=1$. $\Lambda$ plays the role of regulating the deconvolution;
indeed one can see that for Fourier modes present in the image,
$\tilde G({\bmath k})\neq 0$, we have $\lim_{\Lambda\rightarrow 0^+}
\tilde K({\bmath k}) = \tilde \Gamma({\bmath k})/\tilde G({\bmath
  k})$.

To summarise, Eq.~(\ref{eq:kg}) finds the convolving kernel $K$ that
makes the final PSF $G\otimes K$ as close as possible (in the
least-squares sense) to $\Gamma$ without amplifying the noise.  The
kernel is truncated into a $13\times 13$ pixel region centred at the
origin in order to avoid boundary effects and to prevent problems such
as bad columns, saturated stars, or cosmic rays from ``leaking'' all
over the field.  We also re-scale the resulting kernel to integrate to
unity ($\tilde K({\bmath 0})=1$) but since $\Lambda$ is small,
typically of order $10^{-5}$, this has no practical effect.  Note that
since $G({\bmath x})$ and $\Gamma({\bmath x})$ are both real
functions, it follows that in Fourier space they satisfy the
conditions $\tilde G({\bmath k}) = \tilde G^\ast(-{\bmath k})$ and
$\tilde\Gamma({\bmath k}) = \tilde\Gamma^\ast(-{\bmath k})$, and then
Eq.~(\ref{eq:kg}) guarantees that a similar condition holds for $K$:
the convolution kernel $K({\bmath x})$ is real.

The second problem -- the variation of the PSF across the field -- is
handled by taking the reconstructed PSF on a grid of $8\times 6$
points separated by 298 pixels (2 arcminutes) in each direction, and
constructing a grid of 48 kernels $K$.  The kernels are then
interpolated bilinearly between the four nearest grid points, and then the final image
$F({\bmath x})$ is constructed according to \seq{F({\bmath x}) = \int
K_{\bmath x}({\bmath y})I({\bmath x}-{\bmath y}) \,\rmd^2{\bmath x},
\label{eq:f}} where $K_{\bmath x}$ is the kernel reconstructed at
position ${\bmath x}$ in the field.

The convolution kernel will not capture PSF model fluctuations on
scales below 2 arcminutes. Since the SDSS model PSFs are quadratic
functions over the chip, features at the arcminute scale and smaller are not
captured anyway. We show below that, even at $\theta = 1$ arcmin,
the remaining PSF variations not captured by the kernel are very small
compared to the expected shot-noise errors in the two-point statistics
at those scales.

Obviously there will be instances in which the kernel reconstruction
is not good enough.  Therefore a set of cuts must be applied to the
resulting kernels.  In order to construct these cuts, we consider the
Gaussian-weighted moments of the residual $\Gamma-G\otimes K$, i.e.
\seq{M_{\alpha\beta} = \frac{1}{\pi\sigma_1^2}\int [\Gamma-G\otimes
K]({\bmath x})\,\frac{x_1^\alpha
x_2^\beta}{\sigma_1^{\alpha+\beta}}\rme^{-{\bmath
x}^2/2\sigma_1^2}\,\rmd^2{\bmath x}.}  The cuts are then:
\newcounter{cuts}
\begin{list}{\arabic{cuts}. }{\usecounter{cuts}}
\item We reject an entire field if the SDSS software used to determine the
  PSF (the postage stamp pipeline, or {\sc psp}) failed to determine a
good PSF model in the single-epoch imaging, or was forced to use a
low-order fit to the PSF ({\tt PSP\_STATUS!=0}).
\item We reject cases where the PSF residual is too large regardless
of the moments, i.e.  \seq{\frac{\|\Gamma-G\otimes
K\|^2}{\|\Gamma\|^2}>{\tt CUT\_L2}.}
\item We reject cases where the Gaussian-weighted offset is more than
{\tt CUT\_OFFSET} $\sigma_1$, i.e.  \seq{\sqrt{M_{01}^2+M_{10}^2}>{\tt
CUT\_OFFSET}.}
\item We reject cases where the ellipticity of the final PSF exceeds
{\tt CUT\_ELLIP}, i.e.  \seq{\sqrt{(M_{02}-M_{20})^2+(2M_{11})^2}>{\tt
CUT\_ELLIP}.}
\item We reject cases where the PSF size error exceeds {\tt
CUT\_SIZE}, i.e.  \seq{|M_{22}-M_{00}|>{\tt CUT\_SIZE}.}
\item We reject cases where the radial profile of the PSF is severely
in error as determined by the fourth moment, i.e.
\seq{|M_{40}+2M_{22}+M_{04}-2M_{00}|>{\tt CUT\_PROF4}.}
\end{list}

The specific values of the parameters chosen for the cuts depend on
the band and are shown in Table~\ref{tab:params}.  The tightest
constraints on the quality of the PSF are in $g$, $r$, $i$, and $z$
bands ($r$ and $i$ are used to measure galaxy shapes).  In the $u$ band, where the
average image quality is much lower than in the other bands, more
liberal cuts can be applied because we are interested primarily in the
total flux, not the shape.  Also there is more to gain from liberal
cuts because the signal-to-noise ratio in $u$ band is lower.
Nevertheless, a serious error in the size of the PSF will result in
erroneous photometry, and spurious ellipticity could introduce
colour/photo-$z$ or selection biases that depend on galaxy
orientation, so some cuts must be applied.

\subsection{Noise symmetrisation}
\label{sec:c4}

It is a well-known fact in weak lensing that even if the PSF in an
image has been corrected to have perfectly circular concentric
isophotes, it is possible to produce spurious ellipticity if there is
anisotropic correlated noise.  For example, if the PSF is elongated in
the $x_1$ direction and is ``fixed'' by smoothing in the
$x_2$-direction, the resulting map has more correlations in the $x_2$
direction than $x_1$.  This can lead to (1) centroiding biases, in
which the error on the galaxy centroid is larger in the $x_2$ than the
$x_1$ direction, thus yielding more galaxies that appear aligned in
the $x_2$ than $x_1$ direction; and (2) biases in which noise
fluctuations tend to be elongated in the $x_2$ direction, so that
positive noise fluctuations on top of a galaxy (which increase its
likelihood of detection) tend to make it aligned in the $x_2$
direction whereas negative fluctuations (which decrease the likelihood
of detection) make the galaxy aligned in the $x_1$ direction. For a
detailed description of noise-induced ellipticity biases, see
\citet{2000ApJ...537..555K} or \citet{2002AJ....123..583B}. These
phenomena can all mimic lensing signals and hence should be eliminated
from the data.  Our method of doing this is to add synthetic noise to
each field so as to give the noise properties 4-fold rotational
symmetry.  To be precise, we want the power spectrum of the total
noise (actual plus synthetic) to satisfy: \seq{P_N({\bmath k}) =
P_N({\bmath e}_3\times{\bmath k}),\label{eq:4}} where ${\bmath e}_3$
is a vector normal to the plane of the image; the cross product
operation ${\bmath e}_3\times$ rotates a vector by 90 degrees.  Even
though it is not circularly symmetric, this is sufficient to guarantee
zero mean ellipticity for a population of randomly oriented galaxies
because ellipticity reverses sign under 90 degree
rotations.\footnote{In group theory language, the noise properties are
symmetric under the 4-fold rotation group ${\cal C}_4$, which is a
subgroup of the full rotations $SO(2)$.  The condition for zero mean
ellipticity due to noise is that ellipticity fall into one of the
non-trivial representations of the noise symmetry group.}  In principle
$m$-fold symmetry for any integer $m\ge 3$ would suffice, however
4-fold symmetry is the only practical possibility for a camera with
square pixels.  For obvious reasons, we would like to achieve this by
adding the minimal amount of synthetic noise possible.

The simplest way to achieve Eq.~(\ref{eq:4}) is to decompose the power
spectrum into its actual (``act'') and synthetic (``syn'') components:
\seq{P_N({\bmath k}) = P_N^{\rm(act)}({\bmath k}) +
P_N^{\rm(syn)}({\bmath k}).}  The actual component is the white
noise variance $v$ in the input image, smoothed by the convolving
kernel: \seq{P_N^{\rm(act)}({\bmath k}) = v|\tilde K({\bmath k})|^2.}
Since $K$ is real, this power spectrum has 2-fold rotational symmetry:
$P_N^{\rm(act)}({\bmath k}) = P_N^{\rm(act)}(-{\bmath k})$.  The
minimal synthetic noise power spectrum that satisfies Eq.~(\ref{eq:4})
is then \seq{P_N^{\rm(syn)}({\bmath k}) = {\rm max}\left[
P_N^{\rm(act)}({\bmath e}_3\times{\bmath k}) - P_N^{\rm(act)}({\bmath
k}),0 \right].}

Gaussian noise with this spectrum can be obtained by taking its square
root, \seq{\tilde T({\bmath k}) = \sqrt{P_N^{\rm(syn)}({\bmath k})},}
and transforming to configuration space $T({\bmath x})$.  Then one
generates white noise with unit variance and convolves it with $T$.
Since the PSF and hence $K$ varies across the field, $T$ must also
vary; its value is interpolated from the same $8\times 6$ grid of
reference points as used for $K$.

The Gaussian white noise was generated using Numerical Recipes {\tt
gasdev} modified to use the {\tt ran2} uniform deviate generator
\citep{1992nrca.book.....P}.  The seed was chosen by a formula based
on the run, camcol, field number, and filter to guarantee that the
same seed was never used twice in the reductions, and that the same
sequence will be generated if the software is re-run.  
For each field, a sequence of $2048\times 1361$
Gaussian deviates is generated; since there are 128 rows of overlap
between successive fields, we fill in the last 128 rows of each field
with the first $2048\times 128$ deviates from the next field.  It is
also essential that the period of the generator be longer than the
total number of pixels in the survey (of order a few $\times
10^{12}$), a requirement which is {\em not} fulfilled by many
generators, since otherwise the same synthetic ``noise'' pattern will
repeat itself throughout the survey.

The image $F({\bmath x})$ after addition of the synthetic noise is a
{\tt kImage}.

\subsection{Single-image masking}
\label{sec:mask}

Once the kernel-convolved, noise-added image ({\tt kImage}) is
constructed for each run that will contribute to the coadds at a given
position, it is necessary to construct a mask before co-addition.  The
mask must remove the usual image defects as well as diffraction
spikes.  It is constructed as described in this section, and is termed
the {\tt kMask}.

We begin by masking out all pixels in $F({\bmath x})$ for which the
convolution (Eq.~\ref{eq:f}) integrates over a bad pixel.  Since $K$
has compact support -- it is nonzero only in a $13\times 13$ pixel
region -- this means that for each bad pixel in $I({\bmath x})$ we
mask out a $13\times 13$ block in $F({\bmath x})$.  Our definition of
a ``bad pixel'' is one that is out of the field; was interpolated by
{\sc photo} (usually due to being in a bad column); is saturated; is
potentially affected by ghosting (via the {\tt fpM} ghost flag); was not checked for objects by {\sc
photo}; is determined by {\sc photo} to be affected by a cosmic ray;
or had a model subtracted from it.  Note that the first cut means that
a 6-pixel region is rejected around the edge of the field.

The second and more sophisticated mask is applied to remove
diffraction spikes from stars.  The secondary support structure
responsible for the diffraction spikes is on an altitude-azimuth
mount, so that the diffraction spikes appear at position angles of 45,
135, 225, and 315 degrees in the altitude-azimuth coordinate system.
Therefore in the equatorial runs, the orientation of the diffraction
spikes relative to equatorial coordinates changes depending on the
hour angle of observation.  If no correction for this is made, then
after co-addition of many runs, even moderately bright stars have a
hedgehog-like pattern of diffraction spikes at many position angles
that can affect a significant fraction of the area.

Our procedure for removing diffraction spikes is as follows.  We first
identify objects with a PSF flux (i.e. flux defined by a fit to the
PSF) exceeding some threshold (corresponding to $9.7\times 10^5$,
$8.5\times 10^5$, $2.2\times 10^5$, $1.7\times 10^5$, and $1.1\times
10^6$ electrons in filters $r$, $i$, $u$, $z$, and $g$ respectively).
Around these objects, we mask a circle of radius 20 pixels (8 arcsec)
and four rectangles of width 8 pixels (3 arcsec) and length
60 pixels (24 arcsec).  The rectangles have the object centroid at the
centre of their short side, and their long axis extends radially from
the centroid in the direction of the expected diffraction spike.

\subsection{Resampling}
\label{sec:regrid}

In order to co-add images, we must first resample them into a common
pixelization.  Ideally we would like this pixelization to be both
conformal (no local shape distortion) and equal-area (convenient for
total flux measurements).  Unfortunately because the sky is curved, it
is impossible to achieve both of these conditions.  However since our
analysis uses a narrow range of declinations around the equator
($|\delta|\le 1.3^\circ$) we can come very close by choosing a
cylindrical projection; the obvious choices are Mercator (perfectly
conformal) or Lambert (perfectly equal-area).  In our case the
Mercator projection would result in the pixel scale being different by
$\Delta\theta/\theta = 2.6\times 10^{-4}$ at the Equator versus at
$\delta=\pm 1.3^\circ$.  (The area error is twice this, or $5.2\times
10^{-4}$.)  The Lambert projection would preserve shapes at the
Equator but the coordinate system would have a shear of $\gamma =
2.6\times 10^{-4}$ at $\delta=\pm 1.3^\circ$.  Neither of these
problems is particularly serious, since either could if necessary be
corrected in the flux or shape measurements.  We have chosen the
Mercator projection because the small cosmic shear signal means that
we are much more sensitive to a given percentage error in shear than
in flux.  Also, a flux error of $5.2\times 10^{-4}$ is insignificant
compared to the error in the flatfields, so there is no point in
eliminating it at the expense of complicating the shear analysis.

The scale of the resampled pixels must be smaller than the native
pixel scale on the CCD ($\sim 0.396$ arcsec) in order to preserve
information.  However it is desirable for it not to be too small,
since this increases the data volume with no increase in information
content.  It should also not be nearly equal to the CCD scale in order
to avoid production of a moir\'e pattern with large-scale power.  We
have used 0.36 arcsec.

The actual resampling step requires us to interpolate the image from
the native pixelization onto the target Mercator pixelization.  This
is done by 36-point second-order polynomial interpolation on the
$6\times 6$ grid of native pixels surrounding the target pixel\footnote{Polynomial interpolation on an equally-spaced grid of points
converges to sinc interpolation in the limit that the number of
gridpoints is taken to infinity. This is easily seen from the Lagrange interpolation formula
and the infinite product,\\ $\prod_{n=1}^\infty (1-x/n)(1+x/n) = \sin (\pi x)/(\pi x)$.}.  A target pixel is considered
masked if any of the 36 surrounding pixels are masked.

\subsection{Addition of images}
\label{sec:add}

After resampling the images, the next step is to combine them to
produce the co-add.  The combination proceeds in three steps:
comparison of images to reject ``bad'' regions that were not masked in
earlier stages of the analysis; relative sky estimation; and stacking.
Note that bad regions must be explicitly rejected: ``robust''
algorithms such as the median are nonlinear and slightly biased, and result in a final
co-added PSF that depends on object flux and morphology, which is not
acceptable for lensing studies.

Rejection of bad regions is critical because it is possible for some
serious defects such as satellite trails to ``leak through'' earlier
stages of the analysis and not be {\tt kMask}ed.  Rejection at this
stage is also the best way to eliminate solar system objects, most of
which will be known, but which
are not easily identified in the single-epoch {\tt fpC}s but of course
will not show up at the same coordinates in successive runs.  We first
bin each input image into $4\times 4$ resampled pixels.  We then
compare the binned images and reject the brightest or faintest image
if it differs by more than {\tt DELTA\_SKY\_MAX1} from the mean.  When
this rejection is done, we actually mask a $20\times 20$ resampled
pixel region around the affected area.  (We found that without this
padding region, satellite trails were often incompletely masked
because they passed through the corners of some $4\times 4$ regions
and did not sufficiently affect the mean flux.)

Next we compute the difference in sky level among all of the $N$
images.  This difference must be determined and removed because
otherwise a masked pixel in an image with below-average sky will
appear as a bright spot in the co-added image.  We compute the
relative sky level -- an often-neglected step in coaddition -- as follows.  For each pair $(i,j)$ of co-added
images, we compute the difference map $F_i-F_j$ and take the median in
$125\times 125$ resampled pixel blocks.  This is taken as an estimate
of the sky difference $S_i-S_j$.  From these differences we obtain the
unweighted least-squares solution for the sky levels $\{S_i\}$, up to
an additive offset (the absolute sky level cannot be determined by
this procedure).  The mean of these levels is denoted by $\bar
S=\sum_{i=1}^NS_i/N$.  We add to the $i$th image the quantity $\bar
S-S_i$ interpolated to a particular point ${\bmath x}$ by 4-point
interpolation from the nearest block centres.  An entire block is
masked out if $|\bar S-S_i|>${\tt DELTA\_SKY\_MAX2} and if it is an
extremal value (either the highest or lowest sky value).

The stacking of the images works by the usual formula \seq{F_{\rm
    tot}({\bmath x}) = \frac{\sum_{i=1}^N w_i({\bmath x})F_i({\bmath
      x})}{w_{\rm tot}({\bmath x})},} where $w_{\rm tot}({\bmath x}) =
\sum_{i=1}^N w_i({\bmath x})$ and $w_i$ are the weights.  Because the
noise is correlated, the optimal weights are scale-dependent; we have
chosen the optimal weights in the limit of small ${\bmath k}$,
i.e. large scales.  That is, $w_i = v^{-1}$ where $v$ is the noise
variance in image $i$.  For photometry of large objects, $w_{\rm tot}$
can be thought of as an inverse white noise variance, i.e. the mean
square noise flux in a region of area $\Omega$ is $1/w_{\rm
  tot}\Omega$.  However for small objects (which are always our
concern) this is not the case and the error bars must be computed from
the measured noise properties of the co-add.

\begin{figure*}
\includegraphics[width=6.5in]{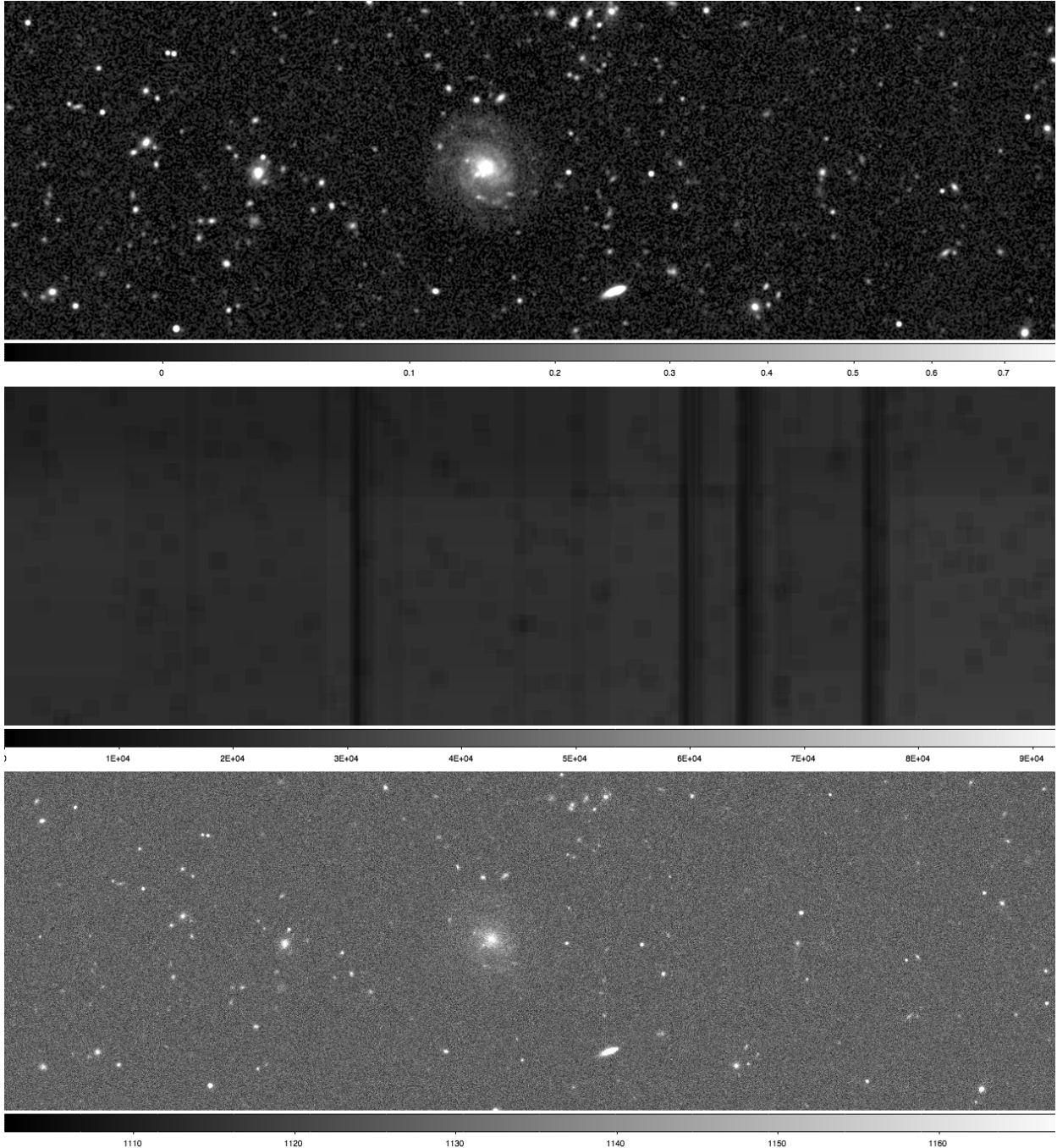}
\caption{\label{fig:singlevscoadd}A comparison of a coadded image (upper panel); its inverse variance map
(middle panel); and a single-epoch input map (bottom panel). The coadded
image is centered on RA $01^h56^m34.8^s$, Dec $-01^\circ10'35''$ (J2000).
East is at top; the image spans $7.7\times 2.4$ arcmin. The top panel
shows the $r$-band image (units: nmgy arcsec$^{-2}$, square root stretch),
and the center panel shows the inverse variance map (units: nmgy$^{-2}$
arcsec$^2$, linear stretch). Note the dark vertical stripes in the inverse variance map produced by bad columns,
and the square patches due to cosmic ray hits propagating through the masking procedure. The spiral galaxy near the center of the image
is of magnitude $r=17.4$. The single epoch image is from strip 82S, run
4263, field 310, camcol 1 (acquired on 2003 November 20 at airmass 1.21).
The image shown is the {\tt fpC} image from rerun 40 on the Data Archive
Server (units: uncalibrated, linear stretch). The same number of pixels are shown, but note
that the single-epoch image is at the native pixel scale (0.396 instead of
0.36 arcsec) and hence shows a slightly larger area.}
\end{figure*}

An example of a co-added image, and comparison to a single-epoch image, is shown in Fig.~\ref{fig:singlevscoadd}.

\subsection{Additional masking}

Before constructing the photometric catalogues, we zero all pixels
contaminated by bright stars in the Tycho catalogues
\citep{2000A&A...355L..27H}, replacing them with random noise of
appropriate amplitude. Pixels masked in this manner have the `INTERP'
bit set in the input fpM files, so that the downstream analysis can
exclude objects that incorporate pixels from a masked region.
Pixels that are {\tt kMask}ed (according to one of the above criteria) also have `INTERP' bits
set. This final step results
in a catalogue with a complex geometry, which will be demonstrated
explicitly in Sec.~\ref{sec:regauss}.

\subsection{Photometric catalogues}

Once each coadded image is constructed, we detect objects using the
catalogue-construction portion of the SDSS photometric pipeline, {\sc
photo-frames}. The details of {\sc frames}'s catalogue construction
and object measurement process are described more fully elsewhere
(\citealt{2002AJ....123..485S}; Lupton et~al., {\em in prep.}).  It is
nevertheless useful to review the important parts of the {\sc frames}
algorithms.

{\sc Photo-frames} requires as input a set of long integer images, and
a considerable array of inputs describing the properties of the
telescope and the observing conditions. Principal among these is a
description of the telescope point-spread function. For single-epoch
data, {\sc frames} uses a principal-component decomposition of the
variation of the PSF across five adjacent fields. The components of
this decomposition are allowed to vary as a polynomial (typically
quadratic) in the image coordinates across each frame. As the coadded
images have the same target PSF in every image, this target PSF is
stored as the first principal component. For fast computation of
object properties, the pipeline uses a double-Gaussian fit to the PSF;
as this is the exact form of the target PSF resulting from the
rounding kernel applied above, we simply use the target PSF
parameters.

{\sc Frames} first smooths the image with the narrower of the two
Gaussian widths describing the PSF. Collections of connected pixels
greater than 7 times the standard deviation of the sky noise are
marked as objects. Each object is grown by six pixels in each
direction. For each object, the list of connected pixels is then
culled of peaks less than three times the local standard deviation of
the sky.

In order to avoid including objects that represent random noise
fluctuations, catalogue galaxies are required to have statistically
significant ($>7\sigma$) detections in both the {\it r} and {\it i}
bands. Note that this is a higher threshold than the $>5\sigma$ cut
used at this stage in the standard single-epoch SDSS processing. This
was necessitated by the fact that the pixel noise in the {\tt kImage}s
is correlated by the convolution process.

In the standard SDSS pipeline, {\sc frames} then re-bins the image and
repeats the search. We choose not to use objects found in this manner,
as the shape measurements of these very low surface brightness
galaxies would not be reliable.

This detection algorithm is repeated in each filter
separately. Objects detected in multiple bands are merged to contain
the union of the pixels in each band if they overlap on the sky. The
list of peak positions in each band is preserved. The centre of the
resulting single object is determined by the location of the highest
peak in the {\it r}\,-band. Objects with multiple peaks are deblended:
the deblending algorithm assigns image flux to each peak in the parent
object.\footnote{Short descriptions of the SDSS deblending can be
found in \citet[][Sec. 4.4.3]{2002AJ....123..485S} and on the SDSS
website at {\tt http://www.sdss.org/dr7/algorithms/deblend.html}. A
detailed paper describing the deblender is forthcoming (Lupton et~al.,
{\it in prep}).}

Once a complete list of deblended peaks (hereafter objects) has been
constructed, the properties of each peak are measured. For the
purposes of this paper, the most important outputs are the {\tt
  MODELFLUX} and {\tt MODELFLUX\_IVAR} parameters\footnote{\tt
  http://www.sdss3.org/dr8/algorithms/magnitudes.php}, which are
determined from the total flux in the best-fit (PSF-convolved) galaxy
profile in the $r$ band (comparing the likelihoods for an exponential
and a de Vaucouleurs model), with the amplitude re-fit separately to
each of the other bands. This flux measure approximates the true,
total flux in the {\it r}\,-band, and provides a robust colour
measurement, which is crucial for photometric estimates of the object
redshift distribution. 

The final crucial output of {\sc photo-frames}, for lensing purposes,
is a postage stamp image for every unique object detected in the
catalogue, except for those objects for which the deblender algorithm
failed.

\subsection{Lensing Catalogue Construction}

After {\sc photo-frames} has constructed an object catalogue from the
coadded images, we attempt to eliminate spurious detections, stars,
and galaxies that are unsuitable for shape measurement. Information
from the input mask (fpM) files is propagated through to the catalogue,
so that objects that incorporate bad pixels identified earlier in the
pipeline can be excluded as needed. Due to the nature of the {\tt
  kImage}s produced by the image coaddition, many of the standard SDSS flags will
not be used (e.g, by construction,
there are no saturated pixels). As we describe above, masked regions of the {\tt kImage}s
are marked as interpolated; objects in the photometric catalogue outputs
with these bits set are removed from the catalogue at this stage. Any
galaxies on which the deblending algorithm failed are also excluded,
as {\sc photo-frames} will not generate unique postage stamps for
these objects.

{\sc photo-frames} also attempts to classify objects as ``stars'' or
``galaxies'' on the basis of the relative fluxes in the point spread
function and galaxy model fits (Lupton et~al., {\it in prep}). Objects
that are well-described by a PSF are classified as stars; we do not
include these objects in the shape catalogue, but set them aside as
aids for detecting systematic errors.

To minimise these effects, we also match against a list of all objects
classified as stars in the single-epoch SDSS catalogues\footnote{As our
sky coverage is less complete than the single-epoch data, we use the
single-epoch catalogues in masking so as to remove objects that are in
close proximity to a star that is in one of our masked regions.}  with
apparent magnitudes in $i$ or $r$ band brighter than 15. We remove
objects from the catalogue within an angular separation of these bright
stars that depends on the stellar apparent magnitude as described in
Table~(\ref{table:starmaskradius}).

\begin{table}
  \caption{Masking radius as a function of apparent stellar magnitude.}
\begin{tabular}{lcc}
  \hline\hline
  Magnitude range & & Masking radius (arcsec) \\
  \hline
  $ r,i\: < 12 $  & & 100      \\
  $12\: < r,i \: < 13 $ & & 70 \\
  $13\: < r,i \: < 14 $ & & 50 \\
  $14\: < r,i \: < 15 $ & & 40 \\
  $15\: < r,i \: < 16 $ & & 30 \\  
  \hline\hline
\label{table:starmaskradius}
\end{tabular}
\end{table}

In addition to these basic cuts, we cull the following objects from the lensing
catalogue:
\begin{enumerate}
\item All objects where the model flux or ellipticity moment
measurement failed;
\item All objects within 62 pixels of the beginning or end of a frame;
\item All objects detected only in the binned images (BINNED2 or
BINNED4);
\item All objects where a bad pixel was was close to the object centre
(INTERP\_CENTER) in either of the $r$ or $i$ bands;
\item All objects that are parents of blends (i.e., measured again in
  terms of the individual child objects);
\item Those for which the observed $r$-band magnitude is greater than 23.5, or the $i$-band magnitude is greater than 22.5.
\end{enumerate}

The magnitude cut was applied to ``observed'' (at the top of the
atmosphere) rather than Galactic extinction-corrected
magnitudes. While this leads to a non-uniform galaxy number density,
it avoids issues with the limiting-$S/N$ varying with position.  
Using the \citet{1998ApJ...500..525S} dust map, with the
standard extinction-to-reddening ratios \citep[Table
22]{2002AJ....123..485S}, along the occupied 100 degree length of the
stripe, the $r$ band extinction $A_r$ has a mean value of 0.141 and a
standard deviation of 0.065. (The $i$ band extinction is lower by a
factor of 0.76.) 
A simple test using the COSMOS Mock Catalogue
\citep[CMC,][]{2009A&A...504..359J} and a size cut\footnote{For an
  $r_{\rm eff}$ of the PSF of 0.67 arcsec and a resolution factor cut
  at $R_2>0.333$, we expect the minimum $r_{\rm eff}$ of a usable
  galaxy to be $0.67\sqrt{0.333/(1-0.333)}$ arcsec. This is of course
  only a very rough estimate, but this application of the CMC provides
  a simple and fast way to estimate the impact of marginal changes in
  survey parameters.} at $r_{\rm eff}>0.47$ arcsec predicts that this
standard deviation should result in a $1\sigma$ variation of $\pm 3$
per cent in the galaxy density and $\pm 1$ per cent in the mean
redshift $\langle z\rangle$. The systematic error introduced by
non-uniform depth, which should be second order in the amplitude of
variations, is expected to be negligible for the purposes of the SDSS
analysis. Note however that this will not be true of future projects. 

Many of these cuts are applied in only one band. The result of this
process is to produce two separate shape catalogues, one for each of the
two shape-measurement bands, so there are a small number of galaxies
which appear in only one of the two catalogues.

\begin{figure}
\includegraphics[width=\columnwidth]{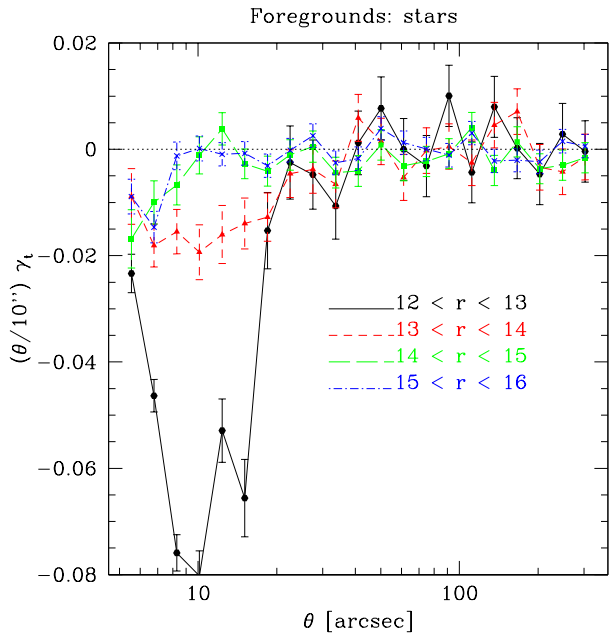}
  \caption{Tangential shear $\gamma_t$ for galaxies as a function of separation
    $\theta$ from stars, as measured in the single epoch SDSS imaging
    using the shape catalogue from \protect\cite{2011arXiv1110.4107R}.  The
    different lines with points show results for bins in $r$-band
    stellar apparent magnitude, as labelled on the plot.  The ideal
    expected value of zero is shown as a dotted horizontal line.}
\label{fig:proximity}
\end{figure}

The SDSS photometric pipeline is known to produce significant sky
proximity effects, wherein the photometric properties of objects
detected near a bright star are systematically biased. The effect of
bright stars on the measured tangential shear of nearby galaxies in
single-epoch SDSS data is shown in
Fig.~\ref{fig:proximity}. Motivated by the scales of the effects
seen there, we mask the regions around bright stars with a masking
radius that depends on the apparent $r$-band (model) magnitude of the
stars as given in Table~\ref{table:starmaskradius}. 

\subsection{Shape measurement}
\label{sec:regauss}

Once the final shape catalogue has been constructed, we use the
re-Gaussianization shape measurement method of
\citet{2003MNRAS.343..459H} to generate an ellipticity measure for
each object. The processing code and script are a modification of
those used in \citet{2005MNRAS.361.1287M}. Re-Gaussianization is not
an especially modern shape measurement technique, but we have used it
previously on SDSS data, it meets our requirements for shear
calibration given the expected statistical power, and we had
well-tested code that interfaced to {\sc photo-frames} outputs at the
time of initiating the cosmic shear project. Therefore we chose to
continue using it for this analysis.

\subsubsection{Overview of re-Gaussianization}

The re-Gaussianization method was an outgrowth of previous work by
\citet{2002AJ....123..583B}. They defined the adaptive moments
${\mathbfss M}_I$ of an image $I$ by finding the Gaussian ${\cal
G}[I]$ that minimises the $L^2$ norm $\| I-{\cal G}[I] \|$. A Gaussian
has 6 parameters -- an amplitude, 2 centroids $\bar{\bmath x}_I$, and
3 components of the symmetric covariance matrix -- and the last of
these {\em is} by definition the $2\times 2$ adaptive moment
matrix. The ellipticity of the galaxy is defined via its components
\begin{equation} e_+^{(f)} = \frac{M_{f,xx} - M_{f,yy}}{ M_{f,xx} +
M_{f,yy}}
\end{equation} and
\begin{equation} e_\times^{(f)} = \frac{2M_{f,xy}}{ M_{f,xx} +
M_{f,yy} }.
\end{equation} For Gaussian PSFs and galaxies, it is easily seen that
the adaptive moment of the intrinsic galaxy image $f$ can be extracted
from that of the observed image via ${\mathbfss M}_f = {\mathbfss M}_I
- {\mathbfss M}_\Gamma$. If the PSF is both circular and Gaussian (a
situation that does not arise in practise) then one can relate the
ellipticity of the observed image to that of the true galaxy image via
the resolution factor $R_{2}$:
\begin{equation} {\bmath e}^{(f)} = \frac{{\bmath e}^{(I)}}{R_{2}}
{\rm ~~and~~} R_{2} = 1 - \frac{T_\Gamma}{T_I},
\label{eq:corr-naive}
\end{equation} where we have used $T$ to denote the trace of the
adaptive moment matrix: e.g., $T_\Gamma\equiv
M_{\Gamma,xx}+M_{\Gamma,yy}$. Re-Gaussianization seeks to apply corrections to
Eq.~(\ref{eq:corr-naive}) to correct for the non-Gaussianity of the
PSF and the galaxy.\footnote{There are also steps in the
\citet{2003MNRAS.343..459H} code that correct for non-circularity of
the PSF. However since the co-add code has already circularised the
PSF, these portions of the code are vestigial and we do not describe
them here.}

\subsubsection{Non-Gaussian galaxies}

First is the non-Gaussian galaxy correction -- i.e. we consider the
case of a Gaussian PSF and non-Gaussian galaxy.  Appendix C of
\citet{2002AJ....123..583B} used a Taylor expansion method to show
that if the galaxy is well-resolved, then in this case
Eq.~(\ref{eq:corr-naive}) could be corrected by using a different
formula for the resolution factor,
\begin{equation} R_2 = 1 - \frac{(1+\beta^{(I)}_{22})
T_\Gamma}{(1-\beta^{(I)}_{22}) T_I},
\end{equation} where $\beta^{(I)}_{22}$ is the radial fourth moment,
\begin{equation} \beta^{(I)}_{22} = \frac{\int (\rho^4-4\rho^2+2)
I({\bmath x}) {\cal G}[I]({\bmath x}) \,\rmd^2{\bmath x}}{2 \int
I({\bmath x}) {\cal G}[I]({\bmath x}) \,\rmd^2{\bmath x}},
\end{equation} where ${\cal G}[I]$ is the adaptive Gaussian and the
rescaled radius $\rho$ is given by
\begin{equation} \rho \equiv \sqrt{({\bmath x} - \bar{\bmath
x}_I)\cdot{\mathbfss M}_I^{-1}({\bmath x}-\bar{\bmath x}_I)}.
\label{eq:bjr}
\end{equation} This is equivalent to an elliptical version of the
$n=4,m=0$ polar shapelet \citep{2003MNRAS.338...35R,
2003MNRAS.338...48R}, and we have $\beta^{(I)}_{22}=0$ for a Gaussian
galaxy (in practise usually $\beta^{(I)}_{22}>0$).

\subsubsection{Non-Gaussian PSF}

Finally we arrive at the correction for the non-Gaussian PSF.  We
construct a Gaussian approximation $G_1$ to the PSF $\Gamma$,
\begin{equation} \Gamma({\bmath x}) \approx G_1({\bmath x}) = {1\over
2\pi\sqrt{\det{\mathbfss M}_{G_1}}} \exp \left( -{1\over 2} {\bmath
x}^T{\mathbfss M}_{G_1}^{-1}{\bmath x} \right).
\label{eq:gdef}
\end{equation} The choice $G_1$ is chosen according to the adaptive
moments of $\Gamma$. The function $G_1$ is determined from the centroid and
covariance, but the amplitude in Eq.~(\ref{eq:gdef}) is chosen to
normalise the Gaussian $G_1$ to integrate to unity.\footnote{The
reason for doing this is that while this increases the overall power
$\int(\epsilon^2)$ of the residual function, it yields
$\int\epsilon=0$, which ensures that for well-resolved objects
(i.e. objects for which the PSF is essentially a $\delta$-function),
the ``correction'' $\epsilon\otimes f_0$ applied by equation
(\ref{eq:iprime}) does not corrupt the image $I$.}

We may then define the residual function $\epsilon({\bmath x}) =
\Gamma({\bmath x}) - G_1({\bmath x})$.  It follows that the measured image
intensity will satisfy $I = \Gamma \otimes f = G_1 \otimes f + \epsilon
\otimes f$, where $\otimes$ represents convolution.  This can be
rearranged to yield:
\begin{equation} G_1 \otimes f = I - \epsilon \otimes f.
\label{eq:gf}
\end{equation} This equation thus allows us to determine the
Gaussian-convolved intrinsic galaxy image $I'$, if we know $f$.  At
first glance this does not appear helpful, since if we knew $f$ it
would be trivial to determine $\Gamma \otimes f$.  However, $f$ appears in
this equation multiplied by (technically, convolved with) a small
correction $\epsilon$, so equation (\ref{eq:gf}) may be reasonably
accurate even if we use an approximate form for $f$.  The simplest
approach is to approximate $f$ as a Gaussian with covariance:
\begin{eqnarray} f_0 &=& {1\over 2\pi\sqrt{\det{\mathbfss M}_f^{(0)}}}
\exp \left( -{1\over 2} {\bmath x}^T[{\mathbfss
M}_f^{(0)}]^{-1}{\bmath x} \right), {\rm ~with} \nonumber \\
{\mathbfss M}_f^{(0)} &=& {\mathbfss M}_I - {\mathbfss M}_\Gamma,
\label{eq:mf0}
\end{eqnarray} where ${\mathbfss M}_I$ and ${\mathbfss M}_\Gamma$ are the
adaptive covariances of the measured object and PSF, respectively.
Then we can define:
\begin{equation} I' \equiv I - \epsilon \otimes f_0 (\approx \Gamma \otimes
f).
\label{eq:iprime}
\end{equation} The adaptive moments of $I'$ can then be computed, and
the PSF correction of Eq.~(\ref{eq:bjr}) can then be applied to
recover the intrinsic ellipticity ${\bmath e}^{(f)}$.

Simple simulations with (noise-free) toy galaxy profiles indicate that
this method has shear calibration errors at the level of a few percent
depending on the galaxy profile, with the worst performance for de
Vaucouleurs profiles at low resolution and high ellipticity
\citep[fig. 5]{2005MNRAS.361.1287M}. Moreover, simulations of SDSS
data based on real galaxy profiles from COSMOS, single-epoch SDSS
PSFs, and realistic noise levels show that the shear calibration
biases are not markedly different under more realistic conditions
\citep{2011arXiv1107.4629M}.  An investigation of the shear
calibration bias for the SDSS cosmic shear sample is presented in
Paper II.

\begin{figure*}
\includegraphics[angle=-90,width=6.6in]{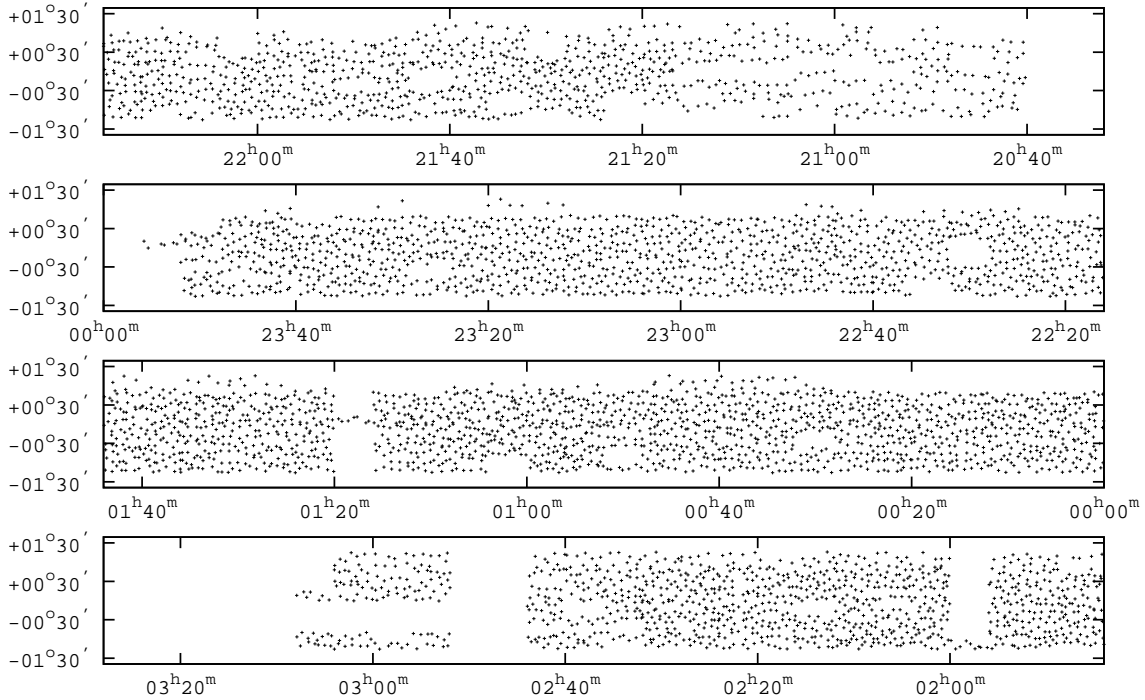}
\caption{\label{fig:gal-i-loc}The angular distribution (in J2000 right
  ascension and declination) of the {\it i}\,-band galaxy catalogue. A
  subsample of every 250th galaxy is shown. The {\it r}\,-band sample
  is identical except for the missing range of
  $-00^\circ48'<$Dec$<-00^\circ24'$. Note the complex survey
  geometry. Coverage gaps at Dec $> 0.8$ are primarily due to the
  severe PSF quality cuts made during the image coaddition step.}
\end{figure*}

\begin{figure*}
\begin{center}
$\begin{array}{c@{\hspace{0.2in}}c}
\includegraphics[width=2.8in,angle=0]{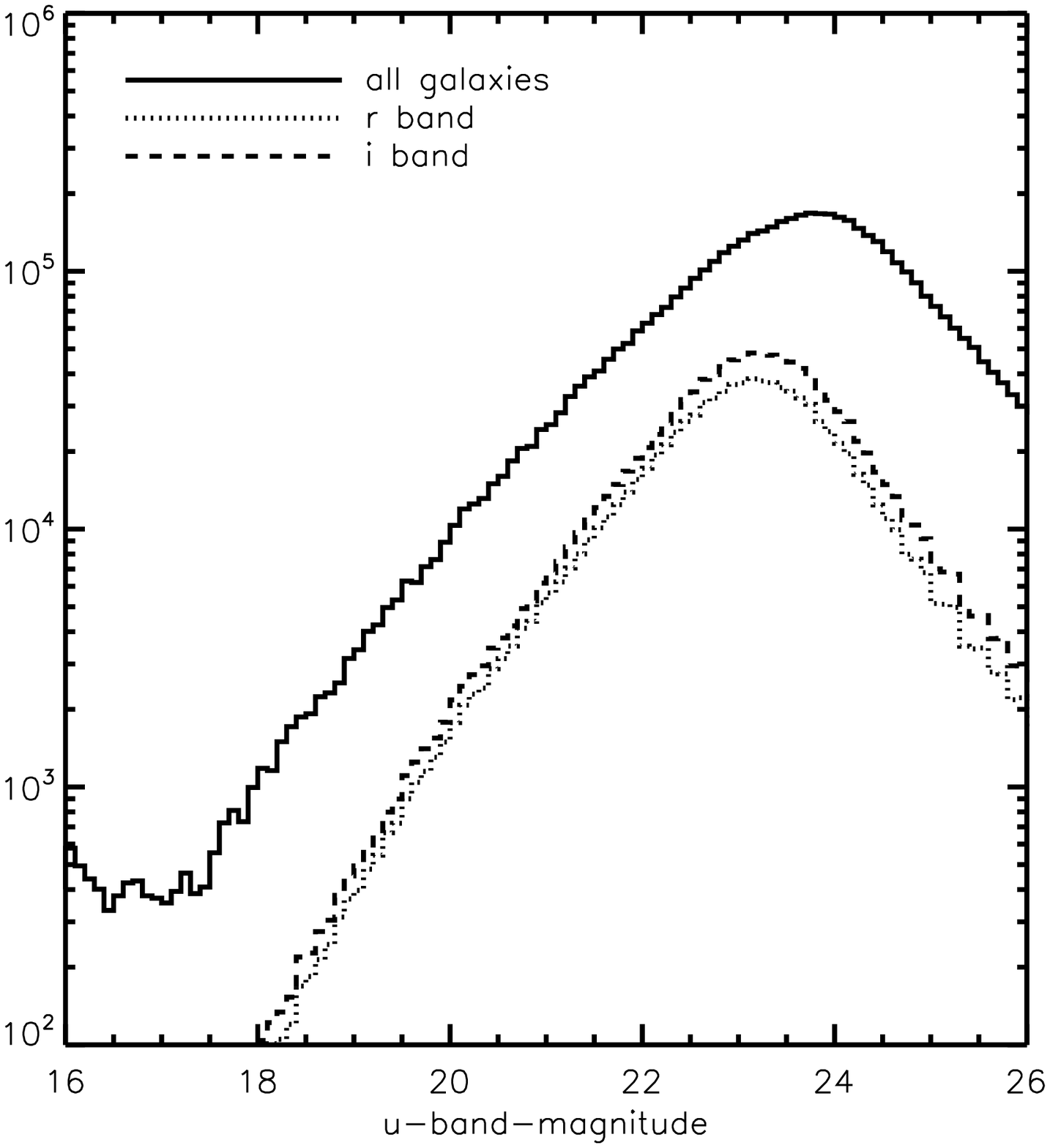} &
\includegraphics[width=2.8in,angle=0]{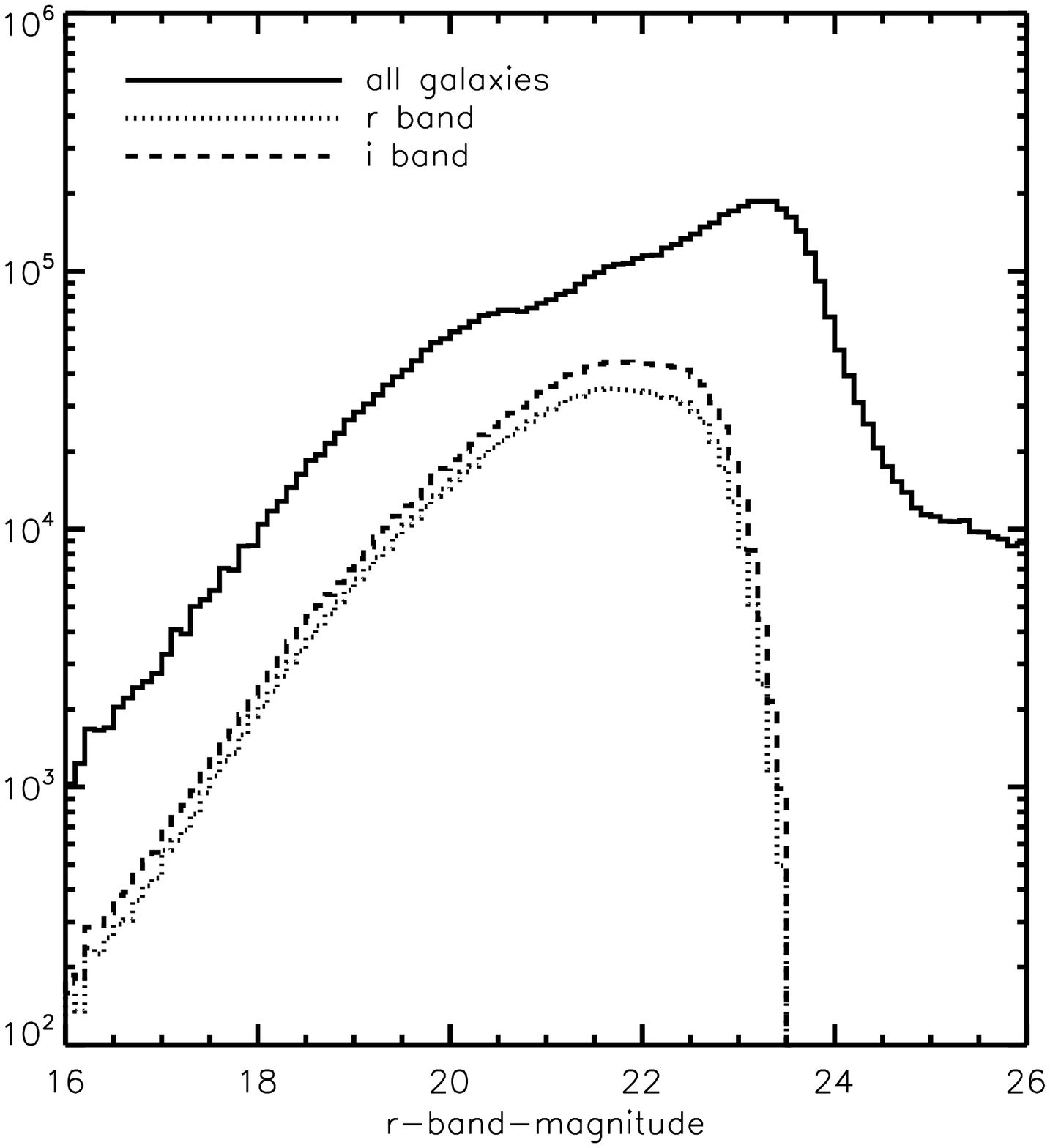} \\
\includegraphics[width=2.8in,angle=0]{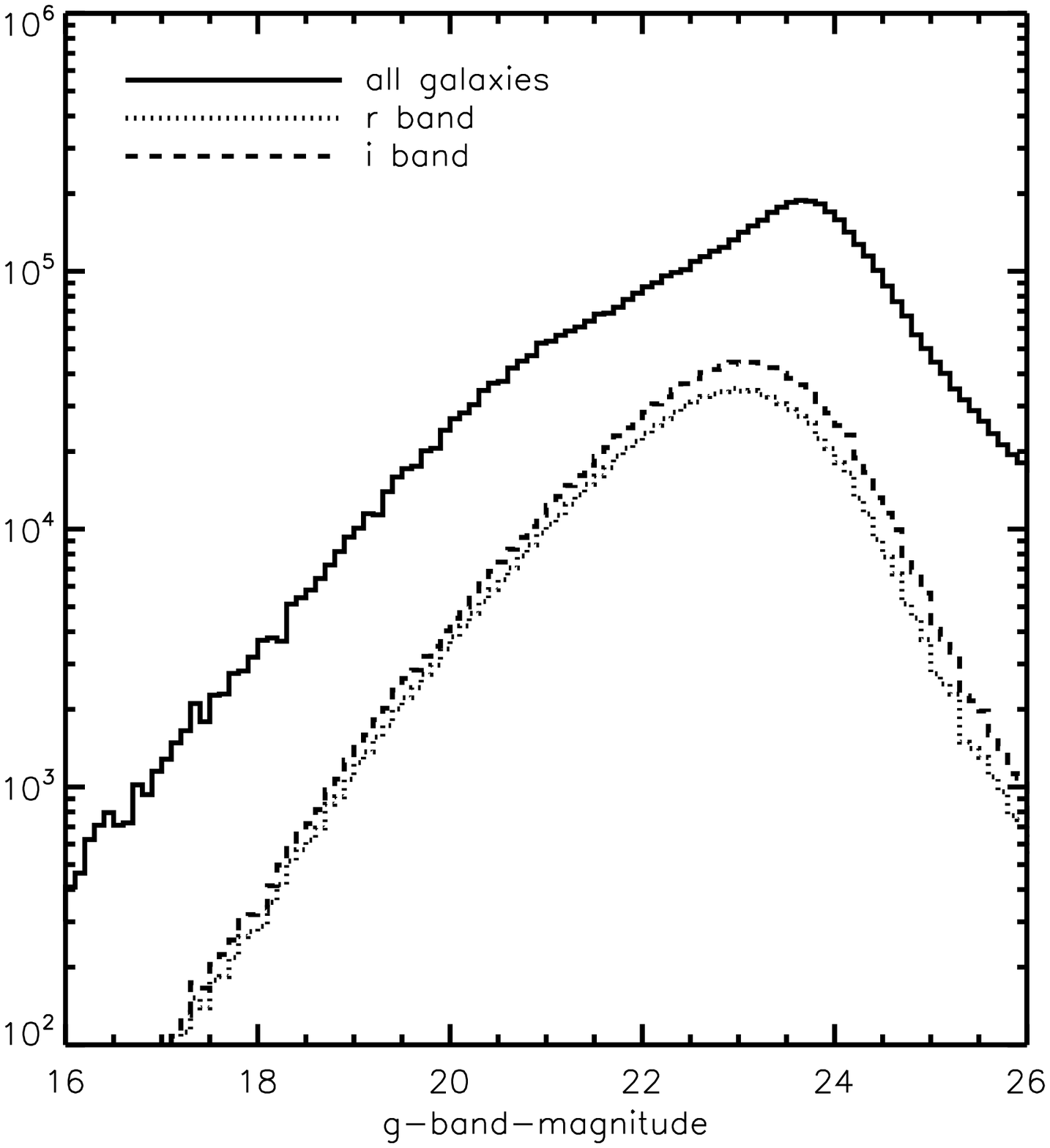} &
\includegraphics[width=2.8in,angle=0]{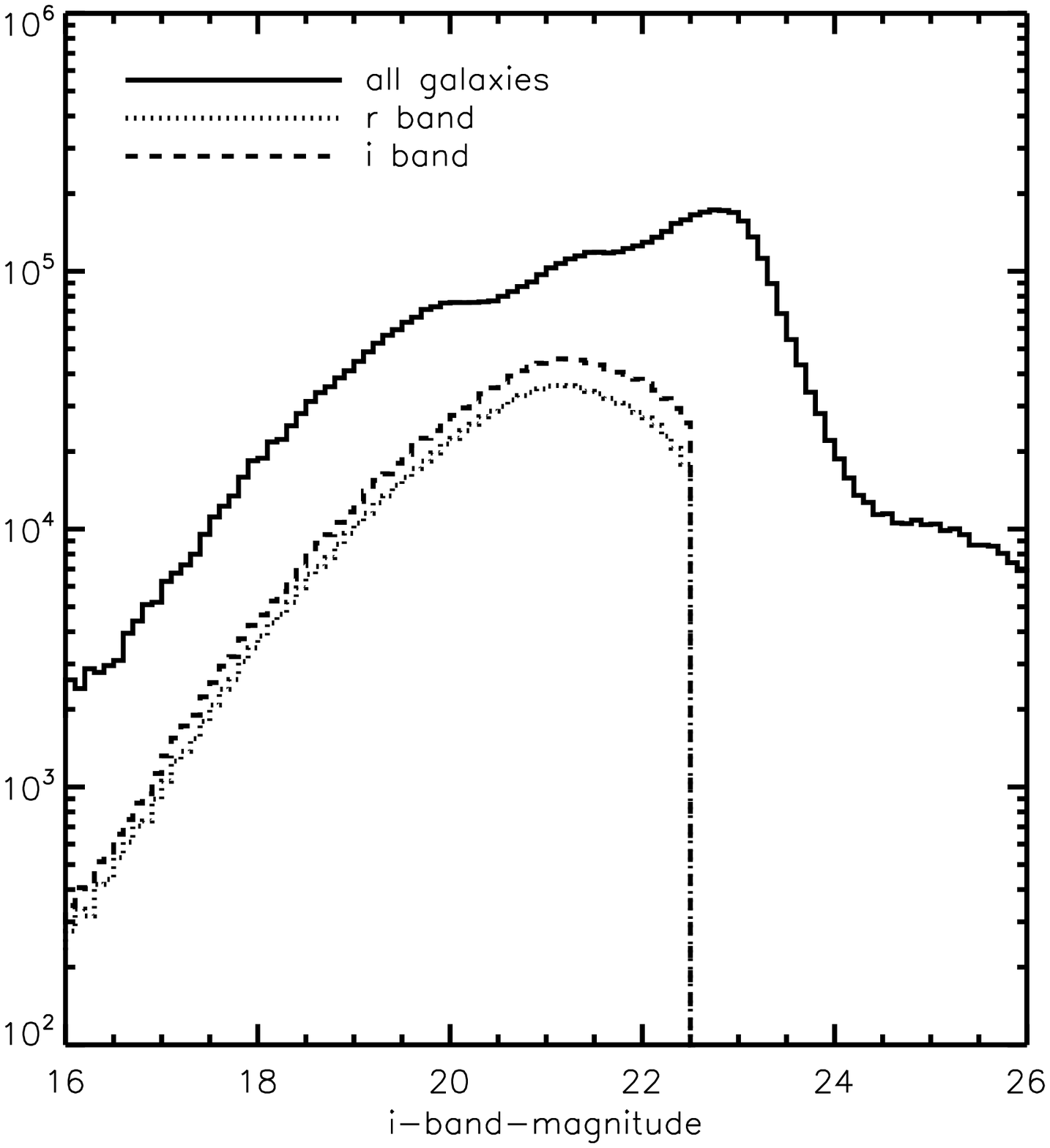} \\
\end{array}$
\includegraphics[width=2.8in,angle=0]{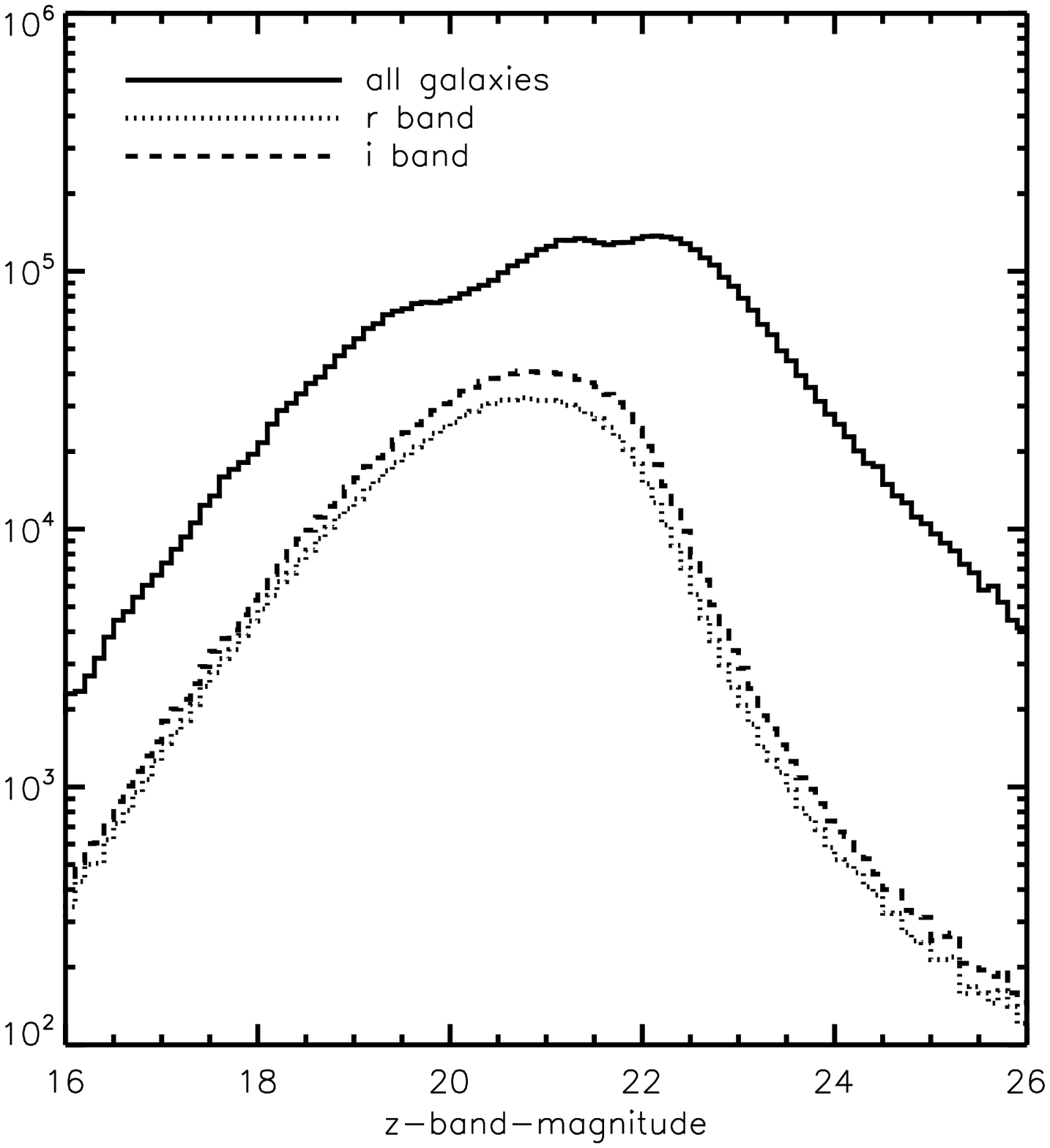}
\caption{\label{fig:mags} The distribution of observed (not
  corrected for Galactic extinction) apparent galaxy model 
  magnitudes in the $u$, $g$, $r$, $i$, and $z$ bands (top left,
  middle left, top right, middle right, and bottom panels).  In all
  cases, the solid line shows the apparent magnitudes for all unique extended objects; dotted and dashed show the $r$- and $i$-band lensing catalogues, respectively.}
\end{center}
\end{figure*}

\begin{figure*}
\begin{center}
$\begin{array}{c@{\hspace{0.2in}}c}
\includegraphics[width=2.8in,angle=0]{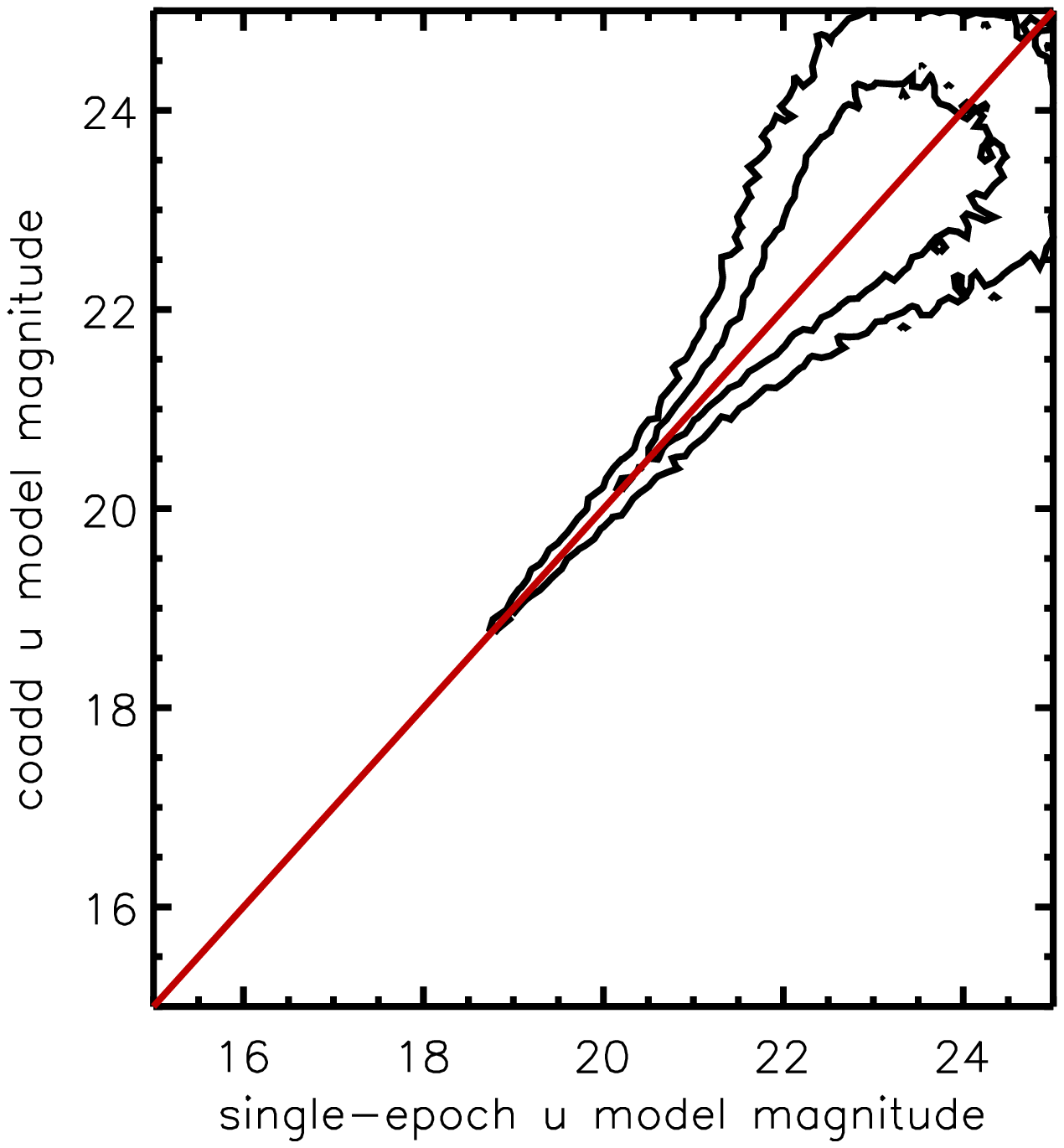} &
\includegraphics[width=2.8in,angle=0]{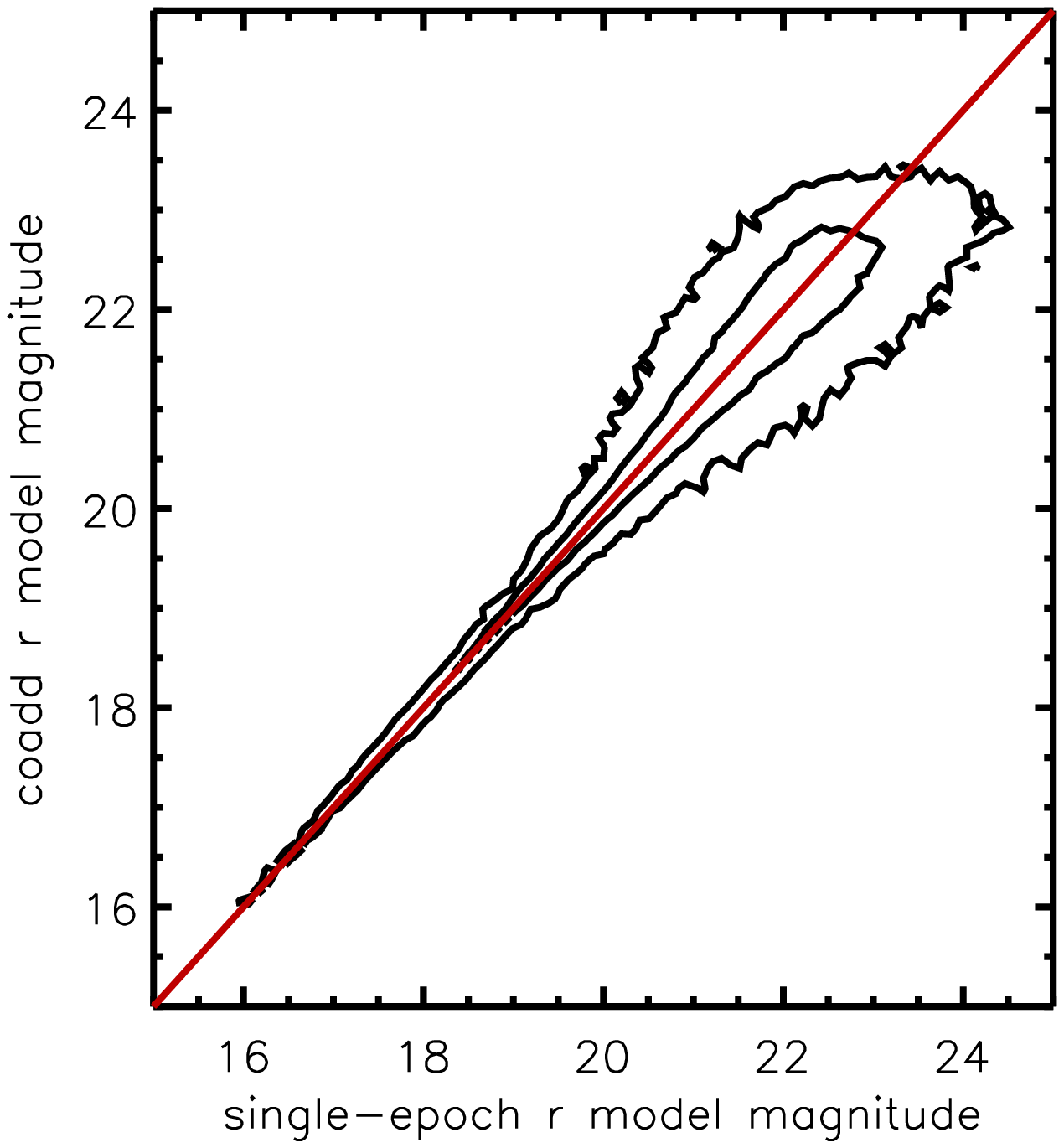} \\
\includegraphics[width=2.8in,angle=0]{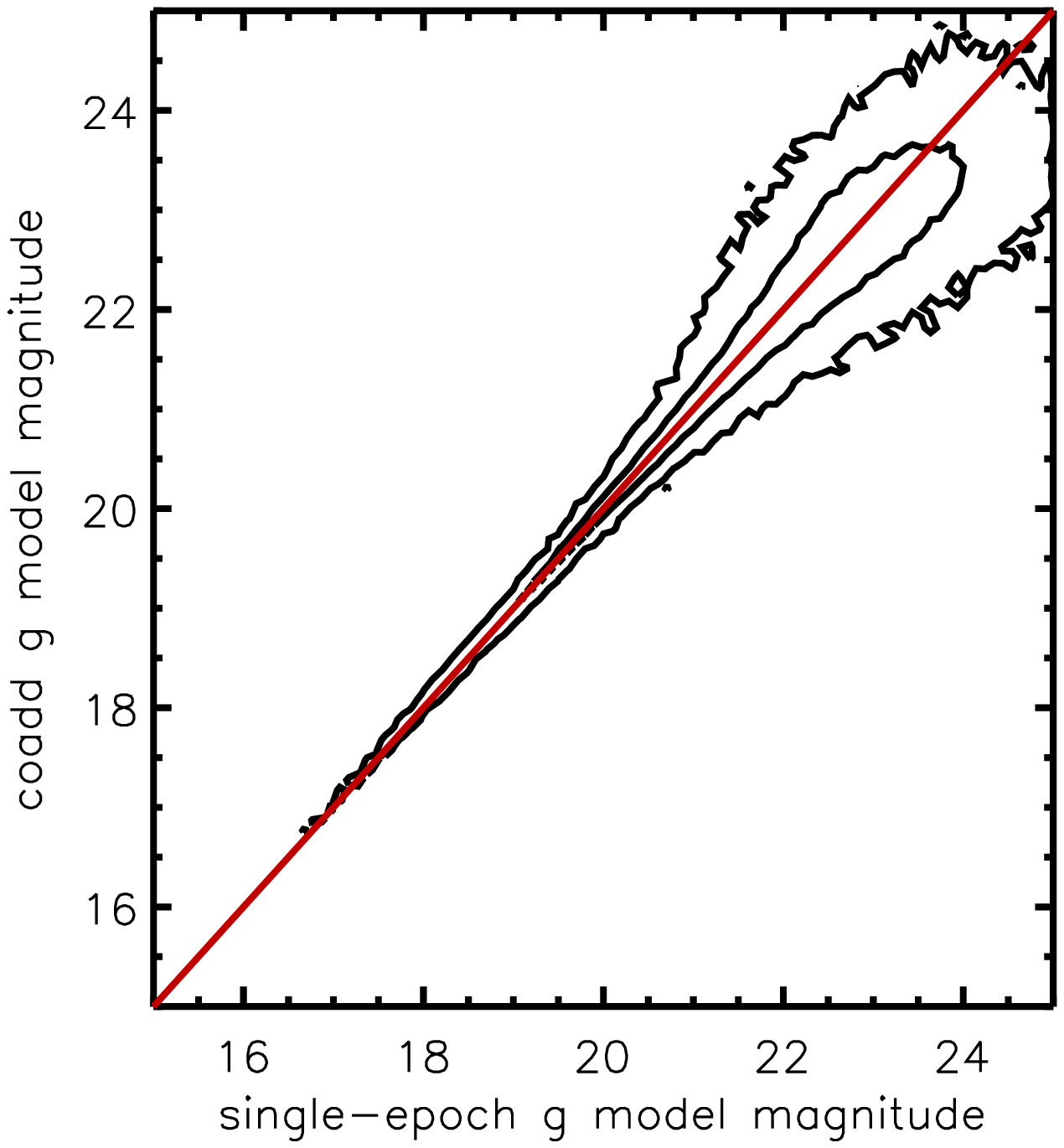} &
\includegraphics[width=2.8in,angle=0]{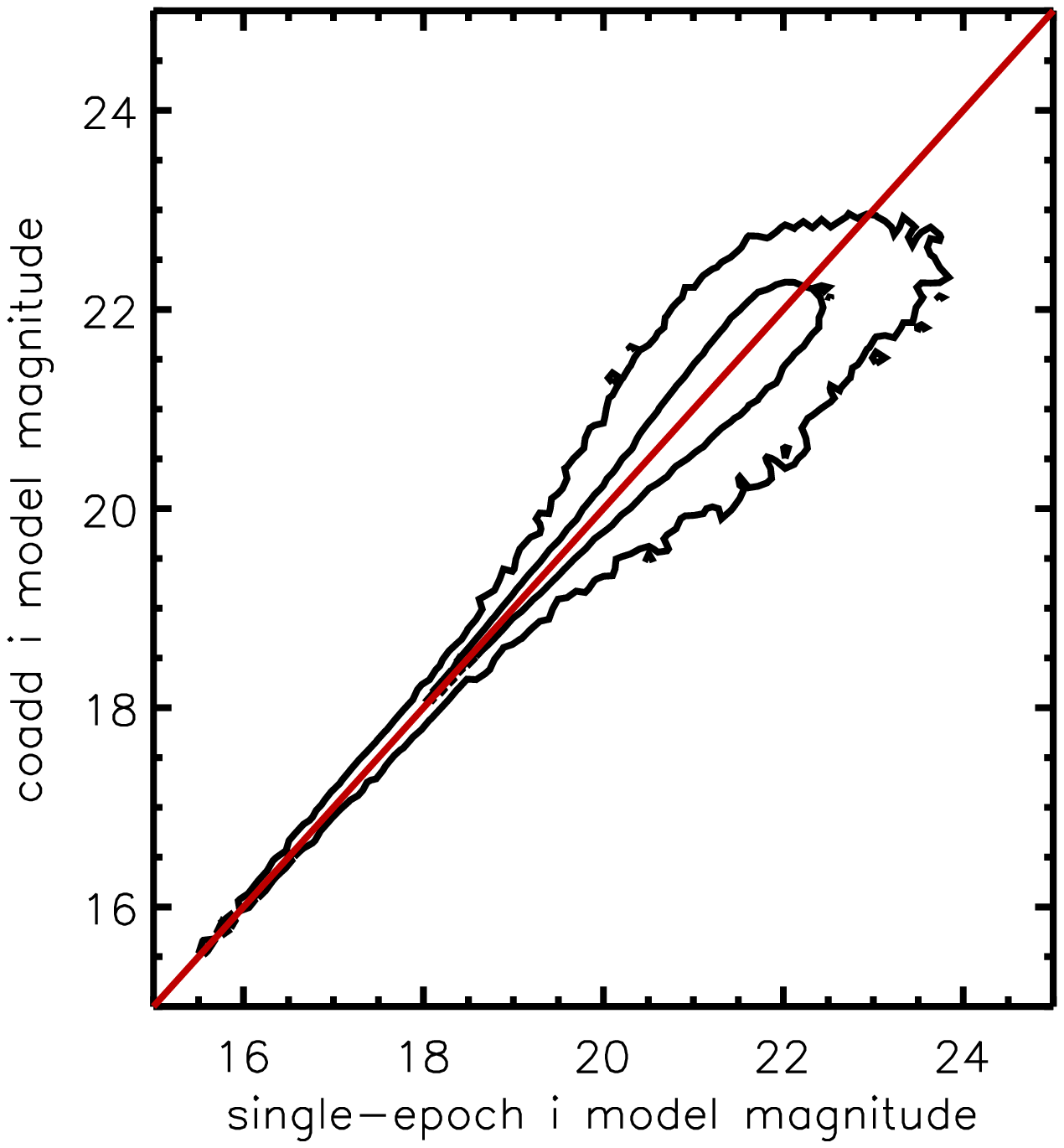} \\
\end{array}$
\includegraphics[width=2.8in,angle=0]{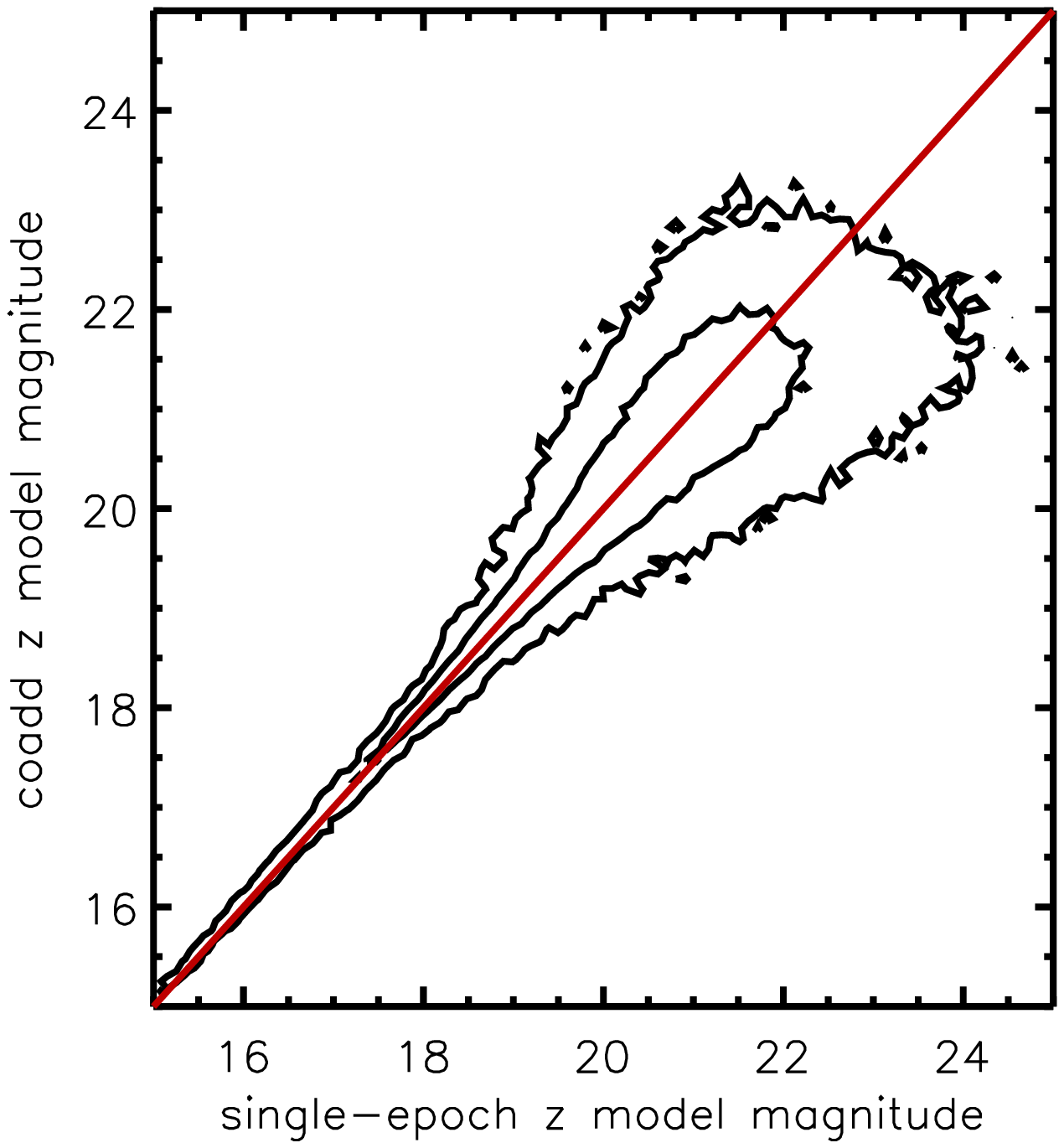}
\caption{\label{fig:compmags} The comparison of the observed (not
  corrected for Galactic extinction) model magnitudes of galaxies in
  the coadd lensing catalogue with magnitudes for the same objects in
  the best run at that position in the single epoch imaging.  Contours
  are 68 and 95 per cent of the total matches.  The asymmetry around
  the 1:1 line at faint magnitudes is due to the flux limit in the
  single epoch images.}
\end{center}
\end{figure*}

To select the galaxy sample used for the final analysis, we impose a
resolution factor cut at $R_2>0.333$ in both $r$ and $i$ (we will
justify this choice in Sec.~\ref{sec:reseffects} based on our desire to minimise
additive PSF systematics).  The
parameters of the final shape catalogue are shown in
Table~\ref{tab:finalshape}, and the survey geometry can be found in
Fig.~\ref{fig:gal-i-loc}.  The apparent magnitude distribution in each
band is shown in Fig.~\ref{fig:mags}.  We show a comparison with the
single-epoch photometry for a representative subsample of galaxies in Fig.~\ref{fig:compmags}.

\begin{table*}
\caption{\label{tab:finalshape}Parameters of the shape
catalogue.}
\begin{tabular}{lcrrcl} 
\hline\hline Parameter & & \multicolumn 2c{Value} & & Units \\ 
& & {\it r}\,-band & {\it i}\,-band & & \\
\hline Total number of source galaxies & & \multicolumn 2c{1\,328\,885} & & ~ \\ 
Number of sources per band & & 1\,067\,031 & 1\,251\,285 & & ~ \\ 
Effective number of sources downweighted by noise, $N_{\rm eff} = \sum_i\varpi_i$ & & 882\,345 & 1\,065\,807 & & ~ \\ 
\hline Median magnitude & & 21.5 & 20.9 & & mag AB \\ 
Median resolution factor $R_2$ & & 0.55 & 0.53 & & ~ \\ 
RMS measured ellipticity per component (noise not subtracted) & & 0.48 & 0.47 & & ~ \\ 
\hline\hline
\end{tabular}
\end{table*}

\section{Correlation function estimation}\label{sec:cf_estimation}

As stated previously, the primary systematic error of concern in this
paper are additive shear systematics, due to PSF ellipticity leaking
into the galaxy shapes even after the PSF correction is carried
out. This concern will drive our choice of diagnostics to use on the
shape catalogues.  There are several possible choices for diagnostics
that we could use:
\begin{enumerate}
\item 1-point statistics of the star and galaxy shapes: For example,
we calculate the mean stellar and galaxy ellipticities in bins of some
chosen size and look for deviations from zero, including coherent
patterns.  We use this diagnostic in Sec.~\ref{subsec:averageshapes}.
\item The tangential shear as a function of scale around random points
\citep[e.g.,][]{2005MNRAS.361.1287M}: If there is some additive
systematic shear, then on scales that are such that we start losing
lens-source pairs off the survey edge, it will show up as a nonzero
tangential shear.  However, this test alone does not tell us much
about the correlations between systematic shears at different points,
and therefore we ignore it in favour of more informative tests.
\item Cross-correlations between the stellar shapes and galaxy shapes,
as a function of separation $\theta$: These correlation functions tell
us not only about the amplitude of any systematic shear, but also
about the characteristic scales that are affected by it.  This section
will describe our methodology for calculating these correlation
functions.
\item The $B$-mode shear, which should be zero due to gravitational
lensing: While this test is an important one as it can signal a
variety of problems with PSF correction, it is not strictly a measure
of additive shear systematics.  Thus, we leave this test for
Paper II, which presents the cosmic shear analysis.
\end{enumerate}

\subsection{The estimator and weighting}

In order to compute the star-galaxy cross-correlations, we employ a
direct pair-count correlation function code.  It is slow ($\sim 3$
hours for $2\times 10^6$ galaxies on a modern laptop) but robust and
well-adapted to the Stripe 82 survey geometry. The code sorts the
galaxies in order of increasing right ascension $\alpha$; stars and
galaxies galaxies are assigned to the range
$-60^\circ<\alpha<+60^\circ$ to avoid unphysical edge effects near
$\alpha=0$. It then loops over all pairs with
$|\alpha_1-\alpha_2|<\theta_{\rm max}$. The usual ellipticity
correlation functions can be computed via summation over galaxies $i$
and stars $j$, e.g.,
\begin{equation}\label{eq:cfestimator} \xi_{11,\rm psf}(\theta) =
  \frac{\sum_{\alpha \beta} w_i e_{\alpha 1} {\tt M\_E1}_{\beta}}{\sum_{\alpha \beta} w_{\alpha}}
\end{equation} and similarly for $\xi_{22,\rm psf}$.  Here $e_{i1}$ is
the PSF-corrected galaxy $e_1$ for galaxy index $i$, and ${\tt
M\_E1}_j$ is the stellar $e_1$ derived from the adaptive moments
described in Sec.~\ref{sec:regauss}.  The sum is over pairs with
separation in the relevant $\theta$ bin, and we weight each pair
according only to the weight associated with the galaxy in each pair:
\begin{equation}
w_i = \frac{1}{0.37^2 + \sigma_e^{2}},
\label{eq:weight}
\end{equation} 
Following \cite{2011arXiv1110.4107R}, we have for
weighting purposes adopted an intrinsic shape noise $e_\mathrm{rms}$
per component of 0.37.  The weight of a galaxy relative to a galaxy
with perfectly measured shape is
\begin{equation} \varpi_i = \frac{w_i}{w(\sigma_e=0)} =
\frac1{1+\sigma_e^2/0.37^2}.
\label{eq:varomega}
\end{equation}

Since the imaging is taken in drift-scan mode, which introduces a
potential preferred direction for PSF distortions, we compute our
diagnostic correlations between the components aligned along ($-e_1$
and $-${\tt M\_E1}) and at $45$ degrees to ($e_2$ and {\tt M\_e2}) the
scan direction.

The code works on a flat sky, i.e. equatorial coordinates
$(\alpha,\delta)$ are approximated as Cartesian coordinates. This is
appropriate in the range considered, $|\delta|<1.274^\circ$, where the
maximum distance distortions are $\frac12\delta^2_{\rm max} =
2.5\times 10^{-4}$. 

All of our shape correlations are computed over the range $1< \theta <
120$ arcminutes, evenly spaced in $\log \theta$.

\subsection{Statistical errors}

The direct pair-count correlation function code can directly compute
the Poisson error bars, i.e. the error bars neglecting the
correlations in $e_{i\alpha}{\tt M\_E\alpha}_{j}$ between different
pairs. This estimate of the error bar is
\begin{equation} \sigma^2[\xi_{++}(\theta)] = \frac{\sum_{i} w_i^2
|{\bmath e}_i|^2 |{\tt M\_Ej}|^2}{2\left( \sum_{i} w_i \right)^2}.
\end{equation} Equivalently this is the variance in the correlation
function that one would estimate if one randomly re-oriented all of
the galaxies. As the star-galaxy correlations described here are
approximate indicators of the amplitude of the additive PSF shear, and
not precision estimates for use in a cosmic shear analysis, we will
not attempt to infer the covariance matrix for the full diagonal
star-galaxy cross-correlation functions.

\section{Diagnostics}\label{sec:diagnostics}

\begin{figure}
\includegraphics[width=0.9\columnwidth]{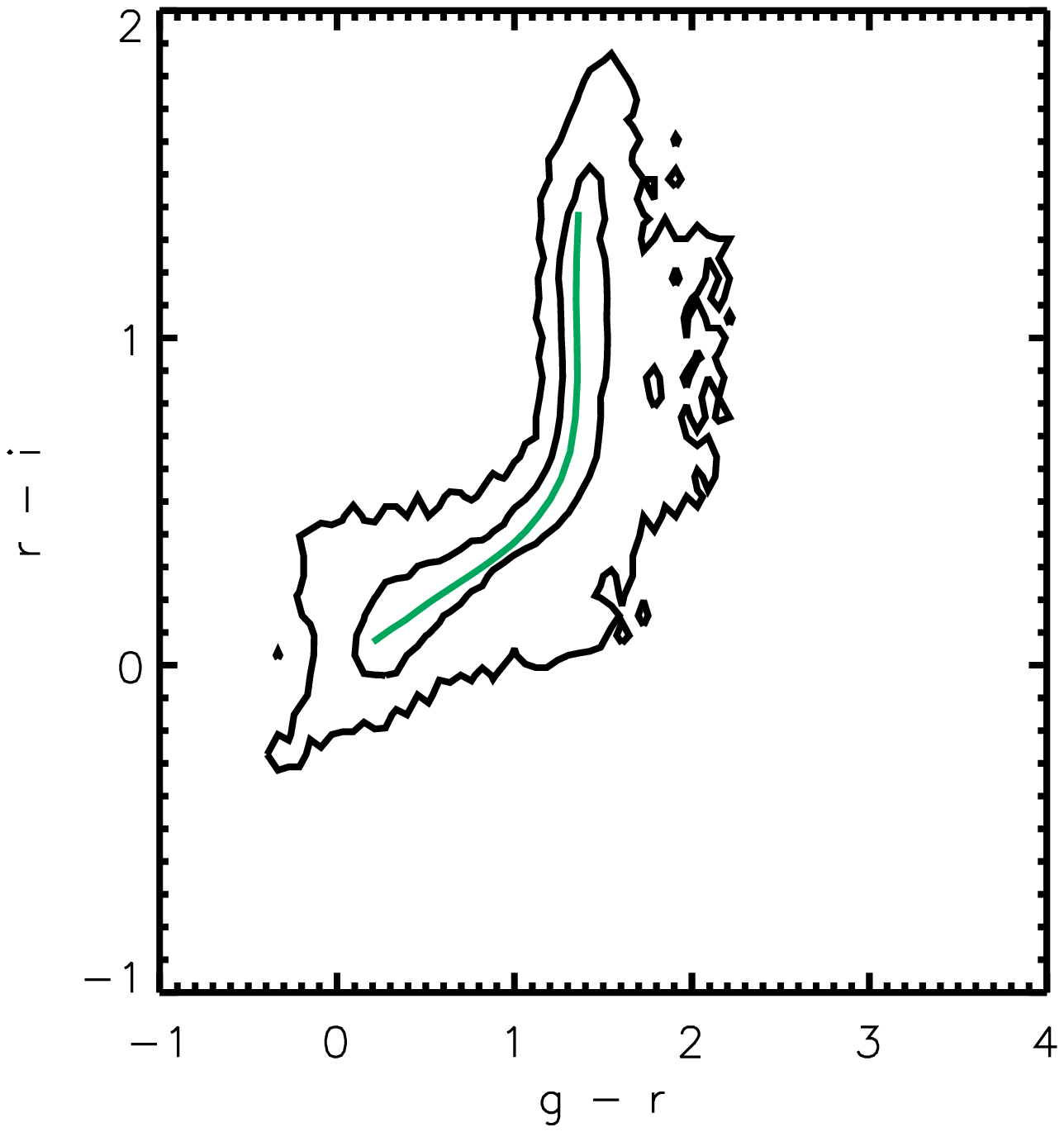}
\includegraphics[width=0.9\columnwidth]{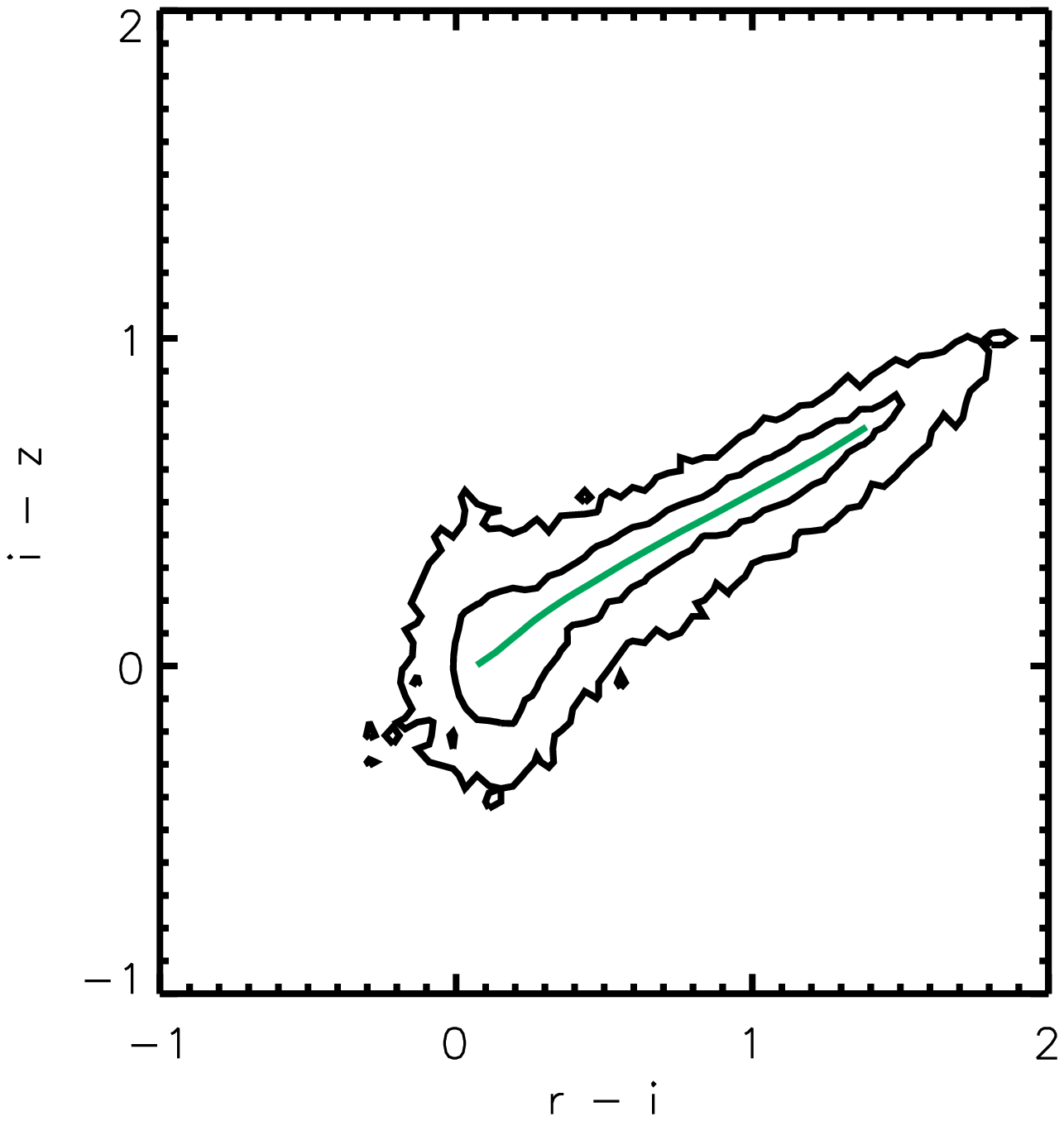}
\caption{Density contour plots in colour-colour space for objects identified as stars using
{\sc Photo}'s star-galaxy separation based on the concentration of the
light profile; the contours containing 68 and 95 per cent of the
density are shown.  The stellar locus from \protect\cite{2002AJ....123.2945R} is shown as a
solid line. This plot includes correction for Galactic extinction,
for fair comparison with previous results.\label{fig:stellar_locus}}
\end{figure}

Here we present our two main systematics tests described in
Sec.~\ref{sec:cf_estimation}, namely the 1-point statistics of the
stellar and galaxy ellipticities, and the star-galaxy shape
cross-correlations.  In order to do this calculation, we must define a
star catalogue, which relies on the {\sc Photo} star-galaxy
separation.  The colours of the objects selected as stars by Photo 
are shown in Fig.~\ref{fig:stellar_locus}. As shown, they agree with
previous determinations of the colours of the stellar locus, e.g.,
from \cite{2002AJ....123.2945R}.

\subsection{Average shapes}\label{subsec:averageshapes}

We first estimate the influence of residual PSF ellipticities on the
galaxy shapes by mapping the stellar shape field.

\begin{figure*}
\includegraphics[angle=-90,width=6.4in]{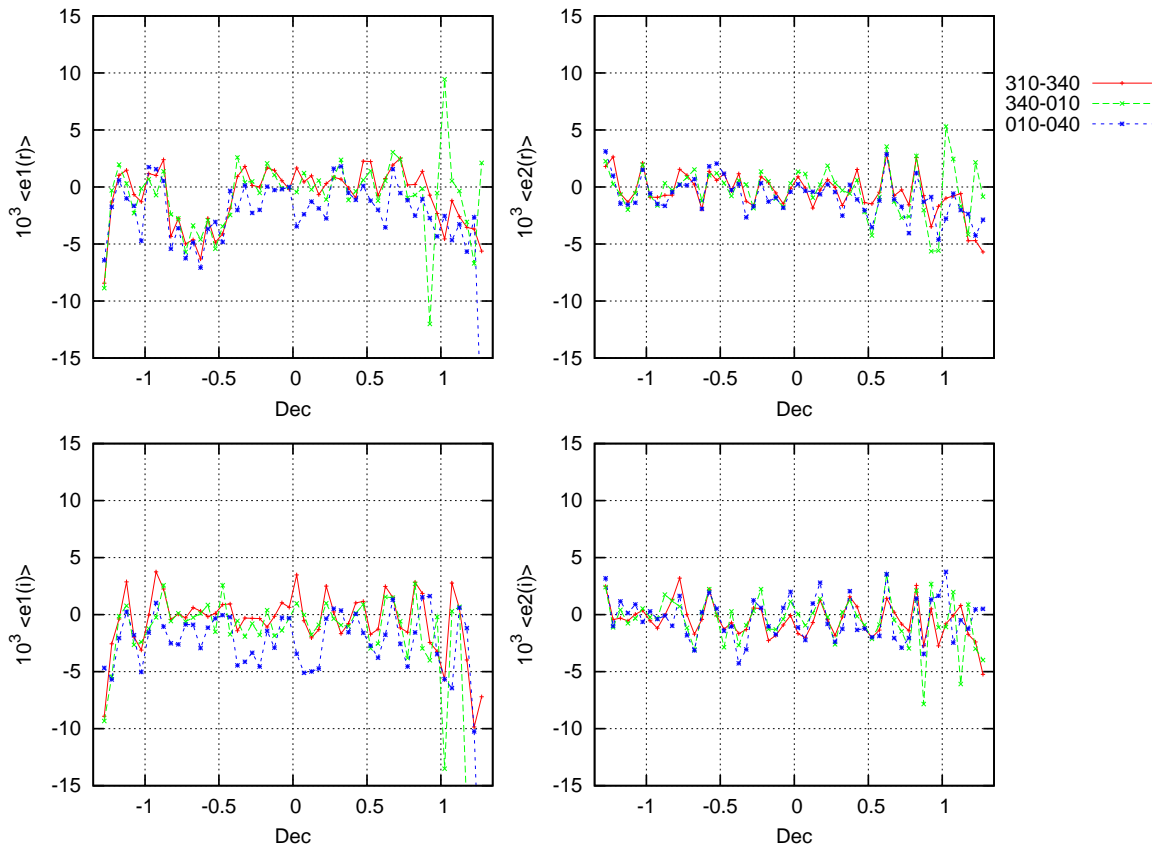}
\caption{The mean ellipticities of stars in the $r$ band as a function
  of declination for different ranges of right ascension, as indicated
  at the upper right. The top
  panels show the $r$ band and the bottom panels show the $i$ band,
  while the left and right panels show different ellipticity
  components. This was computed using a version of the star catalogue
  prior to final cuts. Note the spurious effect in camcol 2 $r$ band
  in the $e_1$ component (declinations $-0.8$ to $-0.4^\circ$). The
  apparent magnitude range for this plot was $19.5<r<21.5 $.}
\label{fig:binnedstars}
\end{figure*}

We computed a set of star shapes binned by right ascension and
declination. The stars were chosen to be moderately faint,
$19.5<r<21.5$, such that they were {\em not} used to estimate the PSF
model in the single-epoch images that was used to construct the
rounding kernel applied to each single epoch image.
Figure~\ref{fig:binnedstars} shows the results: the mean stellar
ellipticities are usually small, of order $10^{-3}$, but in the $r$
band in a particular declination range covered by camcol 2, the shapes
are systematically elongated in the scan direction by $-e_1=0.005$. We
find no significant changes in the amplitude of this artifact when
splitting the stellar populations by colour ($r-i<$ or $>0.3$) or by
apparent magnitude ($r<$ or $>20.5$). We did not definitively
determine the source of this elongation, but we have confirmed that it
appears in the single-epoch SDSS imaging (including the galaxy shape
catalogues from \citealt{2005MNRAS.361.1287M} and
\citealt{2011arXiv1110.4107R}), so is not merely an artifact of the
coaddition and catalogue-making process of this work\footnote{One
  possible explanation is
  incorrect non-linearity corrections for the $r$-band camcol 2
  CCD. The stars used to construct the PSF model are sufficiently
  bright that they require non-linearity corrections, but the stars
  used for our tests here do not. Therefore if the non-linearity
  correction is wrong for that CCD, it could affect the PSF model for
  that CCD alone.}. There is no counterpart feature in the
$i$-band. Given the fact that this feature may plausibly arise due to
problems with the single-epoch PSF model used to determine the proper
convolution kernel to achieve the desired coadd PSF, we exclude all $r$-band galaxy data in camera column 2 from
the cosmic shear analysis.

\subsection{Star-galaxy cross-correlation}

Our primary tasks in producing a shear measurement are to demonstrate
that the additive systematic shear is below the target threshold set
above (Sec.~\ref{sec:design}), and that our shape measurement method
allows us to correctly translate the measured shapes into shears with
sufficient accuracy (a task that we will handle in more detail in
Paper II).

In order to test for residual additive shear systematics, we calculate
the cross-correlation between the measured shapes of the stars and
those of the galaxies in our sample. Any remaining contribution to the
inferred shear field of the galaxies that is sourced by the
point-spread function will produce a non-zero cross-correlation.

\begin{figure*}
\begin{center}
$\begin{array}{c@{\hspace{0.2in}}c}
\includegraphics[width=3in,angle=0]{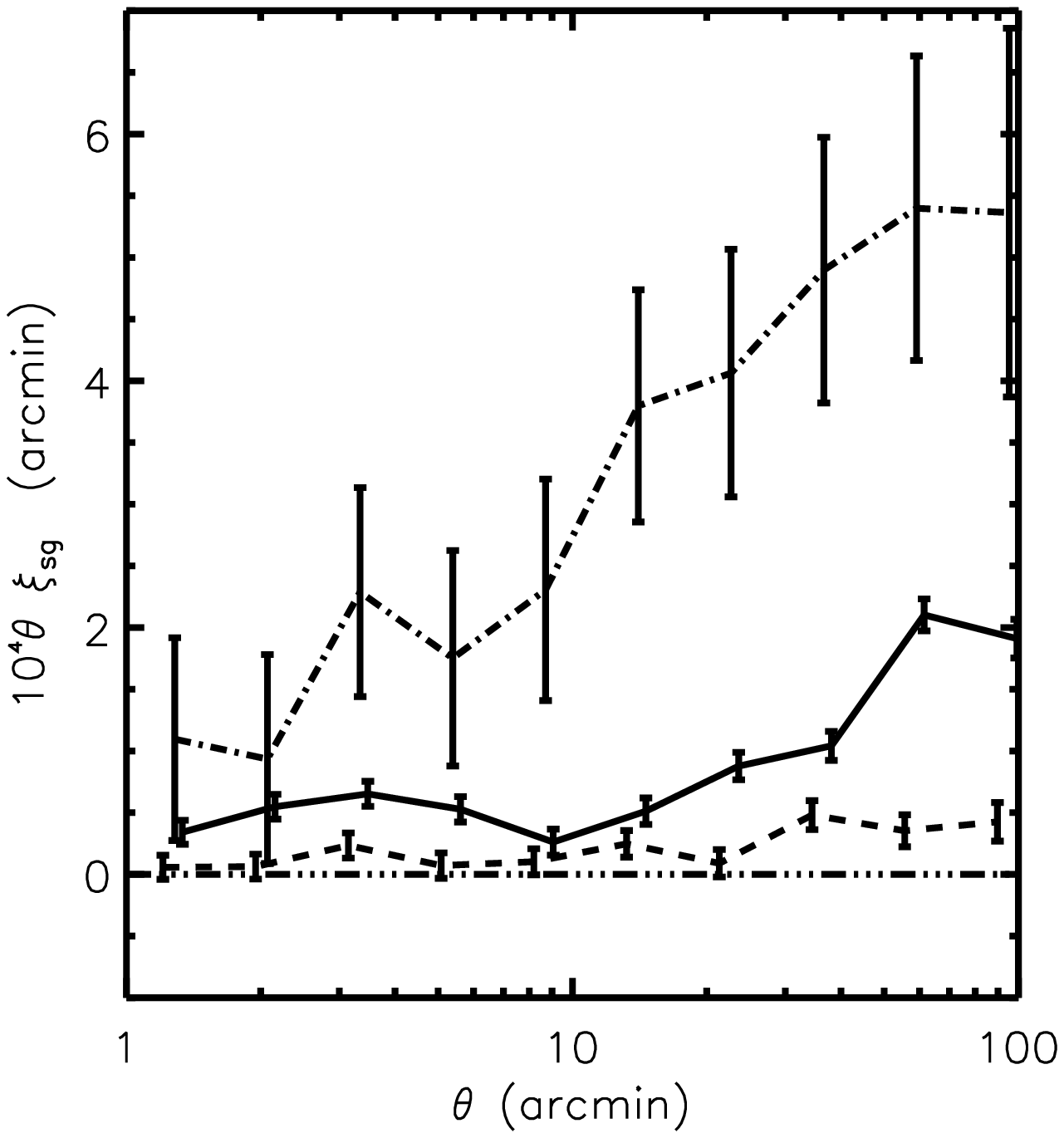} &
\includegraphics[width=3in,angle=0]{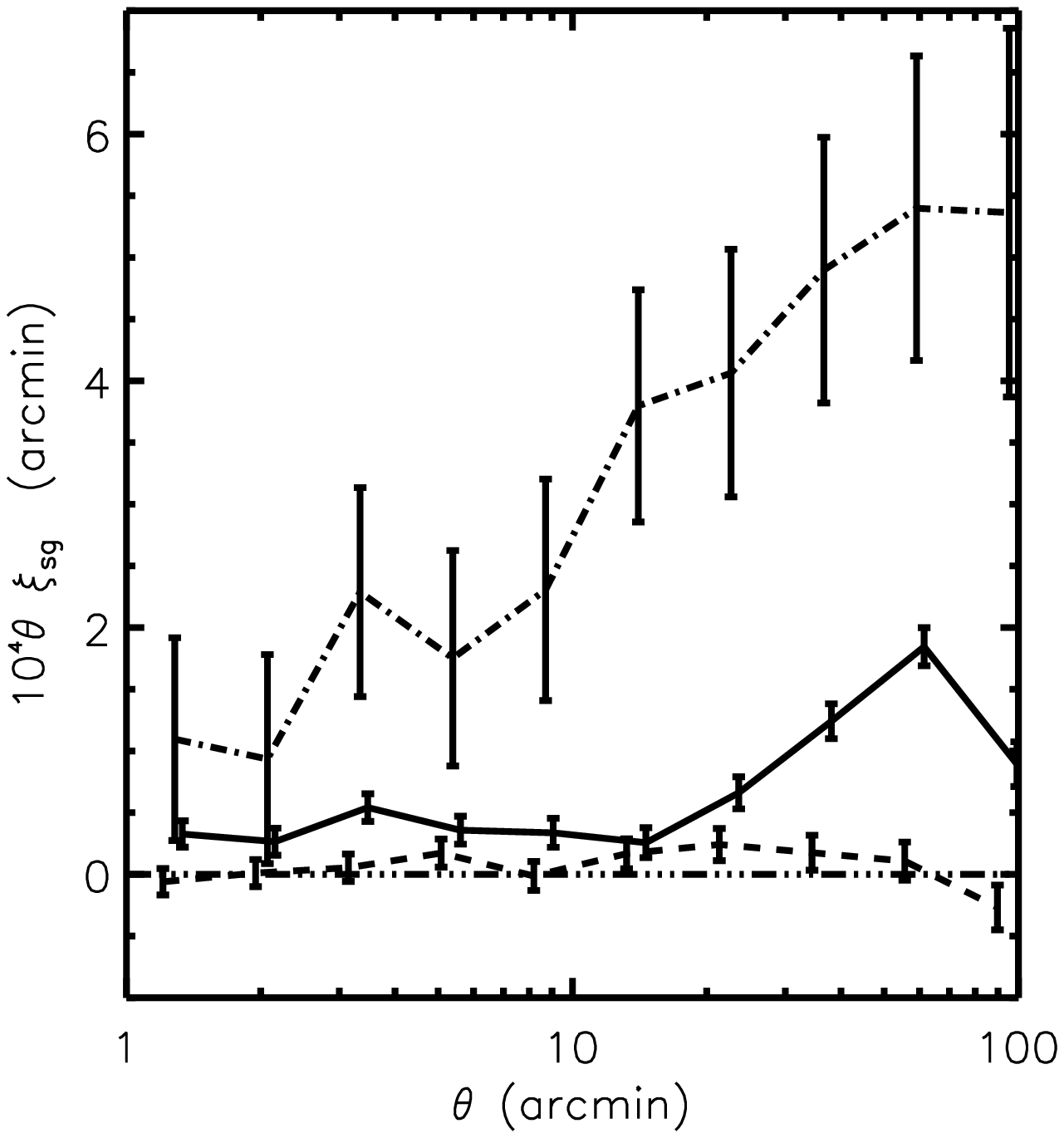} \\
\includegraphics[width=3in,angle=0]{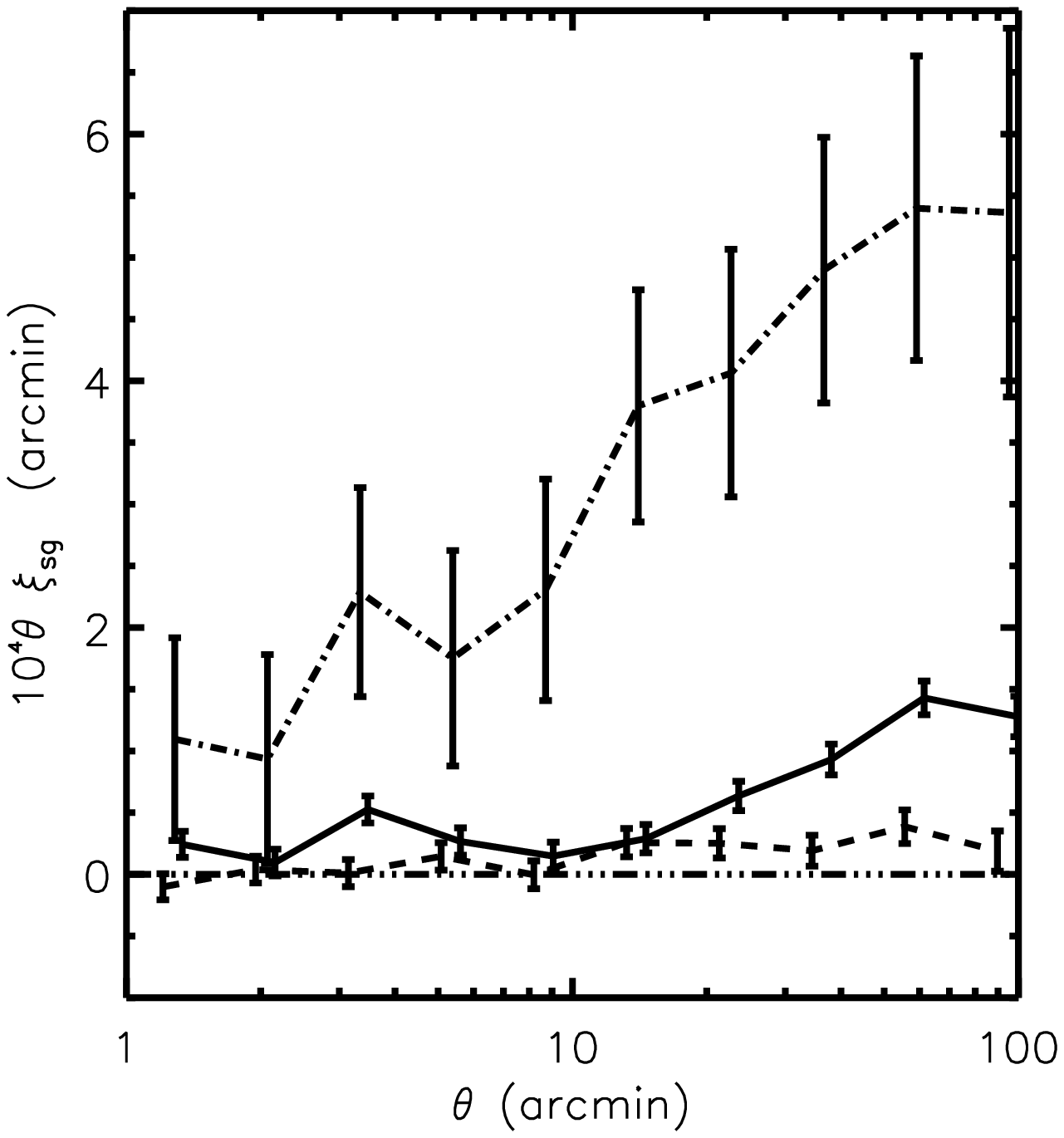} &
\includegraphics[width=3in,angle=0]{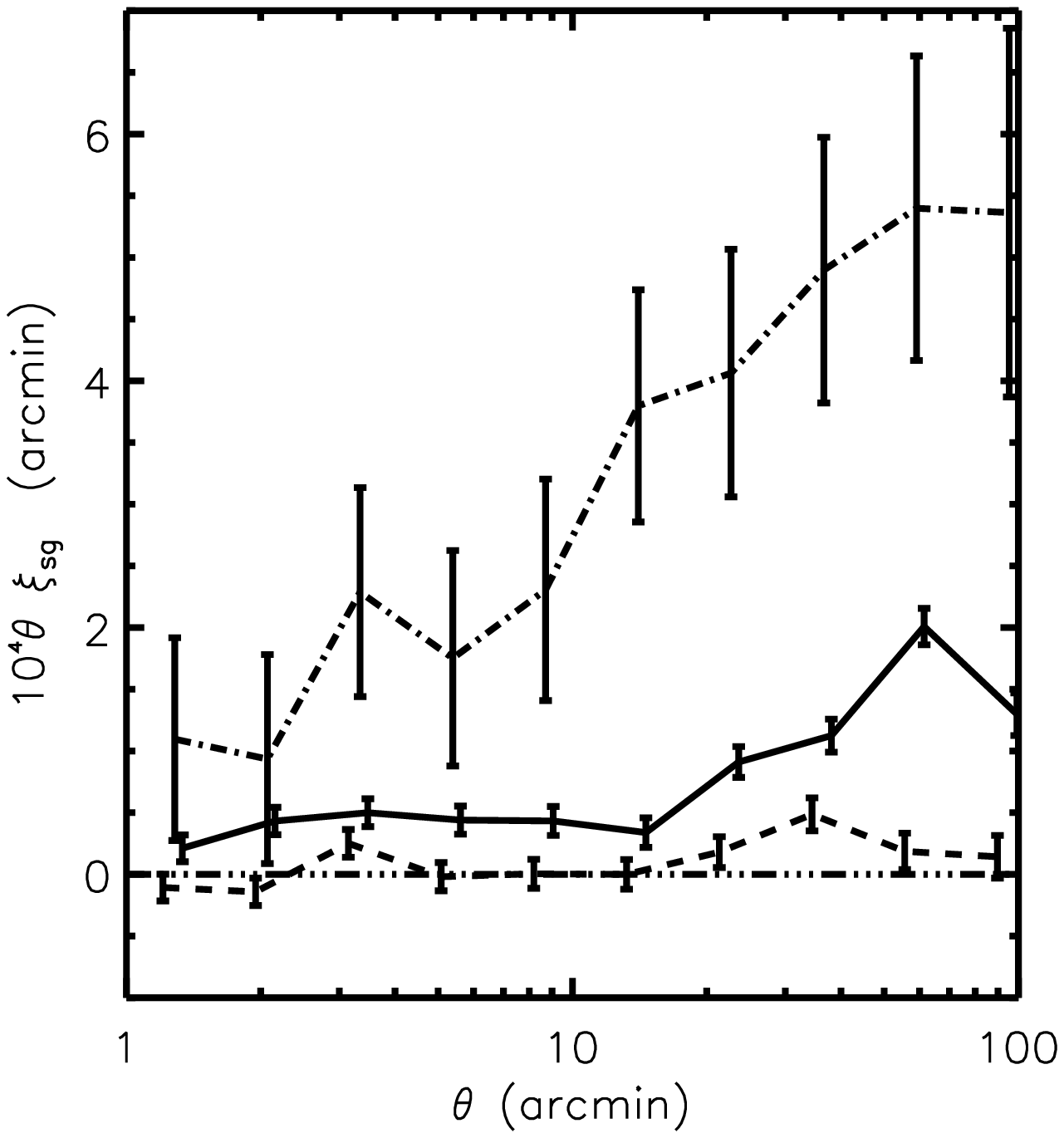} \\
\end{array}$
\caption{\label{fig:sg-all}The cross-correlation of star shapes with
  galaxy shapes, for the following pairs of bands: $(i,i)$ in the
  upper left, $(r,r)$ in the upper right, $(r,i)$ in the bottom left,
  and $(i,r)$ in the bottom right panels. All results are shown as
  $10^{4}\theta\xi$.  The $\left<e_1\,e_1\right>$ correlation is the
  solid line, while the $\left<e_2\,e_2\right>$ correlation is the
  dashed line. The dot-dashed line shows the expected cosmic shear
  $\left<e_+\,e_+\right>$ shape-shape correlation for a survey of this
  depth and size, with shot-noise errors. The triple dot-dashed lines
  shows the ideal value of zero for the star-galaxy
  correlations. Statistics shown are for stars with apparent $i$ and
  $r$ band magnitudes between $19.5$ and $21.5$.}
\end{center}
\end{figure*}

We estimate the star-star and star-galaxy cross correlations as in
Eq.~(\ref{eq:cfestimator}) for all star-galaxy pairs within and
between the $r$ and $i$ bands. The results for the star-galaxy
correlations are shown in Fig~\ref{fig:sg-all}.
For the systematic error diagnostics considered here, we are primarily
interested in computing the cross-correlation between resolved
galaxies and unresolved point sources.

\subsection{Resolution cuts}\label{sec:reseffects}

Due to the PSF dilution correction applied to all galaxy shapes in
Sec.~\ref{sec:regauss}, noisy measurements of poorly resolved galaxies
can significantly amplify any residual additive shear systematics not
corrected for in the rounding kernel process. To assess the effects of
a resolution cut, we compute the star-galaxy cross-correlations in
each band for $R_2>0.25$, $>0.333$, and $>0.4$. Adopting the second of
these of these thresholds appears to be sufficient to minimise the
amplitude of the star-galaxy shape correlation signal. As a result, we
adopt a cut of $R_2>0.333$ for both the {\it i} and {\it r}\,-band
galaxy catalogues.

\subsection{Star-galaxy separation}

\subsubsection{Contamination of star sample by galaxies}

A nonzero amplitude of $\xi_{sg}$ can also be produced by imperfect
star-galaxy separation. Poorly-resolved galaxies masquerading as stars
sample both the PSF- and cosmic shear-sourced shape fields. If the
fraction of stars that are actually mistakenly classified as galaxies
is $f_\mathrm{gal}$, then the measured $\xi_{sg}$ will include a
contribution proportional to $f_{\mathrm gal}\xi_{\gamma}$. As the
ellipticity of nearly-unresolved galaxies will be diluted by PSF
convolution, this represents an upper limit to the level of
star-galaxy correlation that can be introduced via imperfect
star-galaxy separation.

\begin{figure}
\includegraphics[width=3.2in]{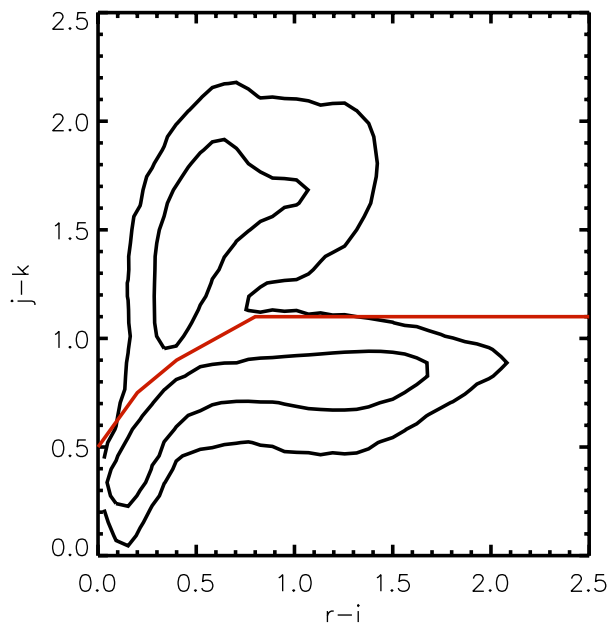}
\caption{\label{fig:ukidss_sep}The cut in $(J-K,r-i)$ space, defined
  using extinction-corrected magnitudes, that was used to separate
stars from galaxies using the UKIDSS data. Objects below the curve (i.e. blue in $J-K$)
are colour-classified as stars, while those above the curve are colour-classified as galaxies.}
\end{figure}

The {\sc photo-frames} pipeline classifies an object as a star or a
galaxy on the basis of the relative fluxes of PSF and galaxy model
fits to the object's surface brightness profile. We have already
confirmed that we get a reasonable stellar locus from this
determination, compared with that from single-epoch imaging
(Fig.~\ref{fig:stellar_locus}).  As another 
check on this scheme, we have defined a sample of stars for which
aperture-matched UKIRT Infrared Deep Sky Survey (UKIDDS) colours are
available.  The UKIDSS project is defined in
\cite{2007MNRAS.379.1599L}. UKIDSS uses the UKIRT Wide Field Camera
(WFCAM; \citealt{2007A&A...467..777C}).  The photometric system is
described in \cite{2006MNRAS.367..454H}, and the calibration is
described in \cite{2009MNRAS.394..675H}. The pipeline processing and
science archive are described in \cite{2008MNRAS.384..637H}.  
Stars and galaxies separate fairly cleanly in $J-K, r-i$ colour space \citep[e.g.,][]{2010MNRAS.404...86B},
so we attempt to use a matched catalogue from Bundy et al. {\em in
  prep.} to put some limits on
galactic contamination of the stellar sample (see
Fig.~\ref{fig:ukidss_sep}). This constraint on
$f_{\rm gal}$ will give us our upper limit $f_{\rm gal}\xi_\gamma$ on
the $\xi_{sg}$ due to contamination of the star sample by galaxies.

We match the objects classified as stars in both bands from our coadd
to UKIDSS objects with valid $J-K$ colours; objects with angular
separations between the two catalogues less than one arcsecond are
identified. We find 93\,753 such stars (as classified by {\sc Photo}). Of these, 11\,331, or 12 per
cent, have $J-K, r-i$ colours inconsistent with the stellar
population. The UKIDSS matches are shallower than the rest of the
catalogue in the $i$ band, but of comparable depth in the $r$ band. Only
16 per cent of our stars have UKIDSS matches in either band, however,
so the contamination fraction is not well-constrained in the entire
star sample.

If, however, this fraction is representative of the galaxy
contamination in the entire stellar catalogue, then for an unresolved
population with a typical resolution just below our resolution cut,
that level of contamination would explain a substantial fraction of the residual PSF
systematic amplitude that we see.

\begin{figure*}
\begin{center}
$\begin{array}{c@{\hspace{0.2in}}c}
\includegraphics[width=3in,angle=0]{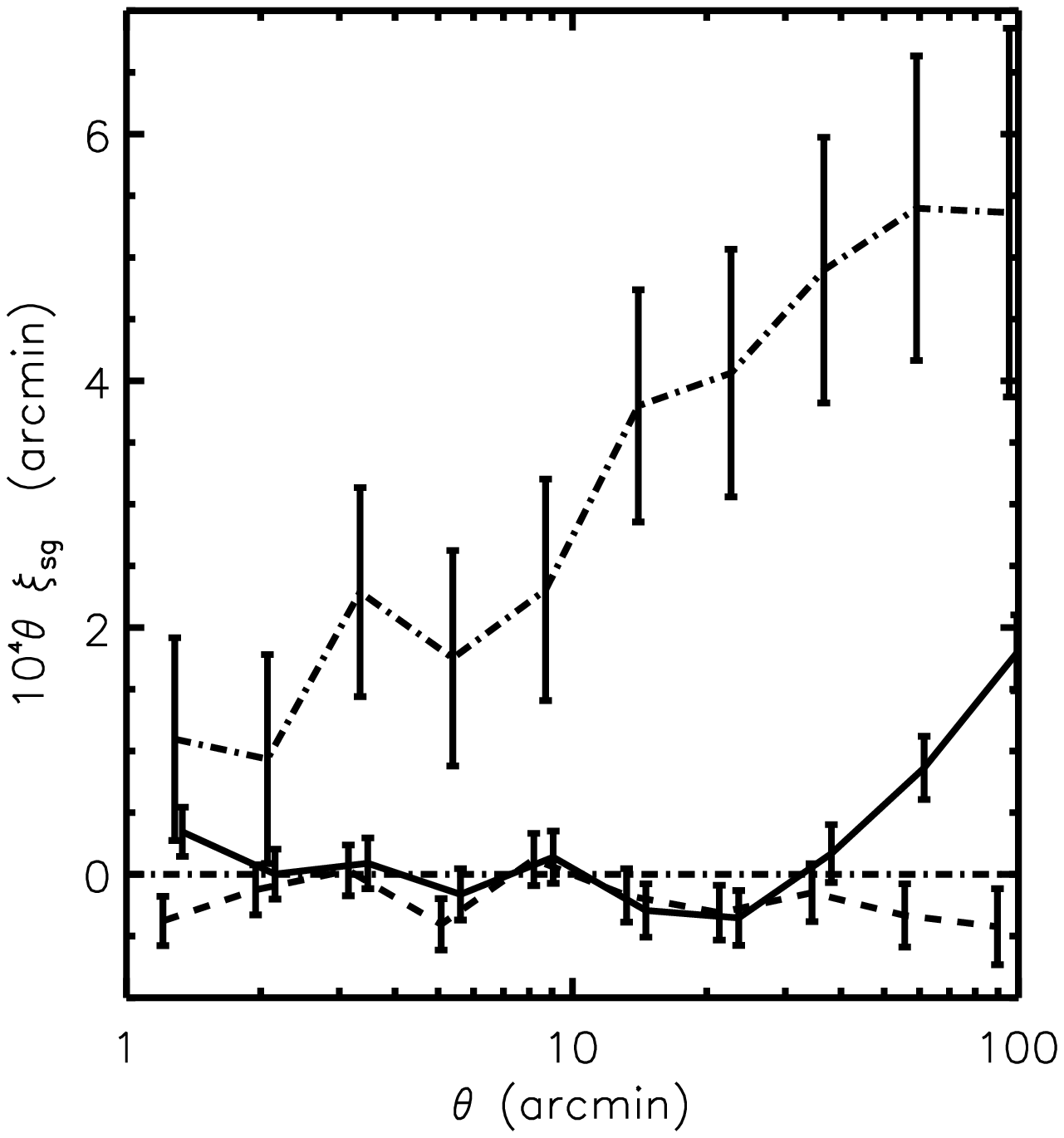} &
\includegraphics[width=3in,angle=0]{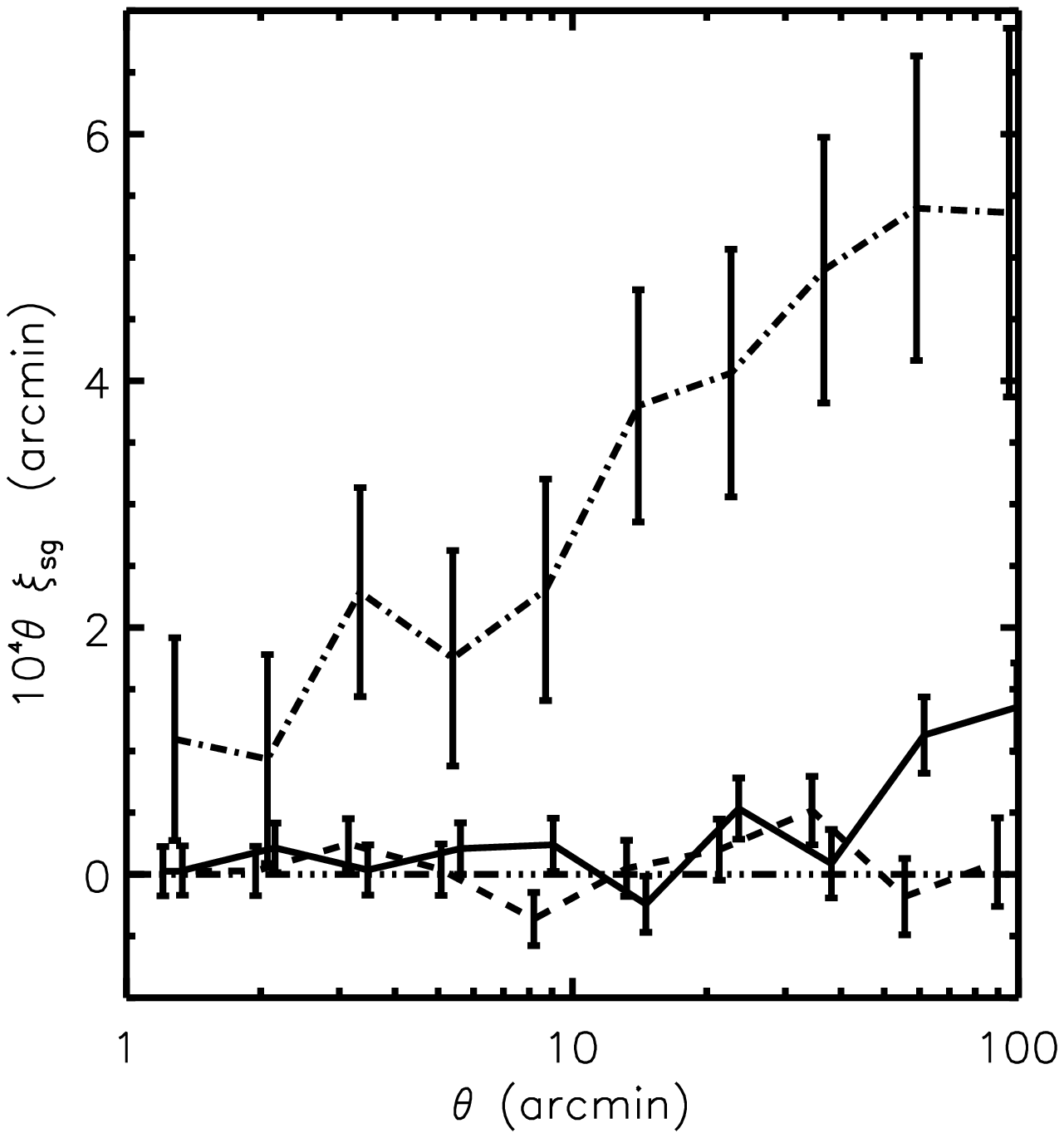} \\
\includegraphics[width=3in,angle=0]{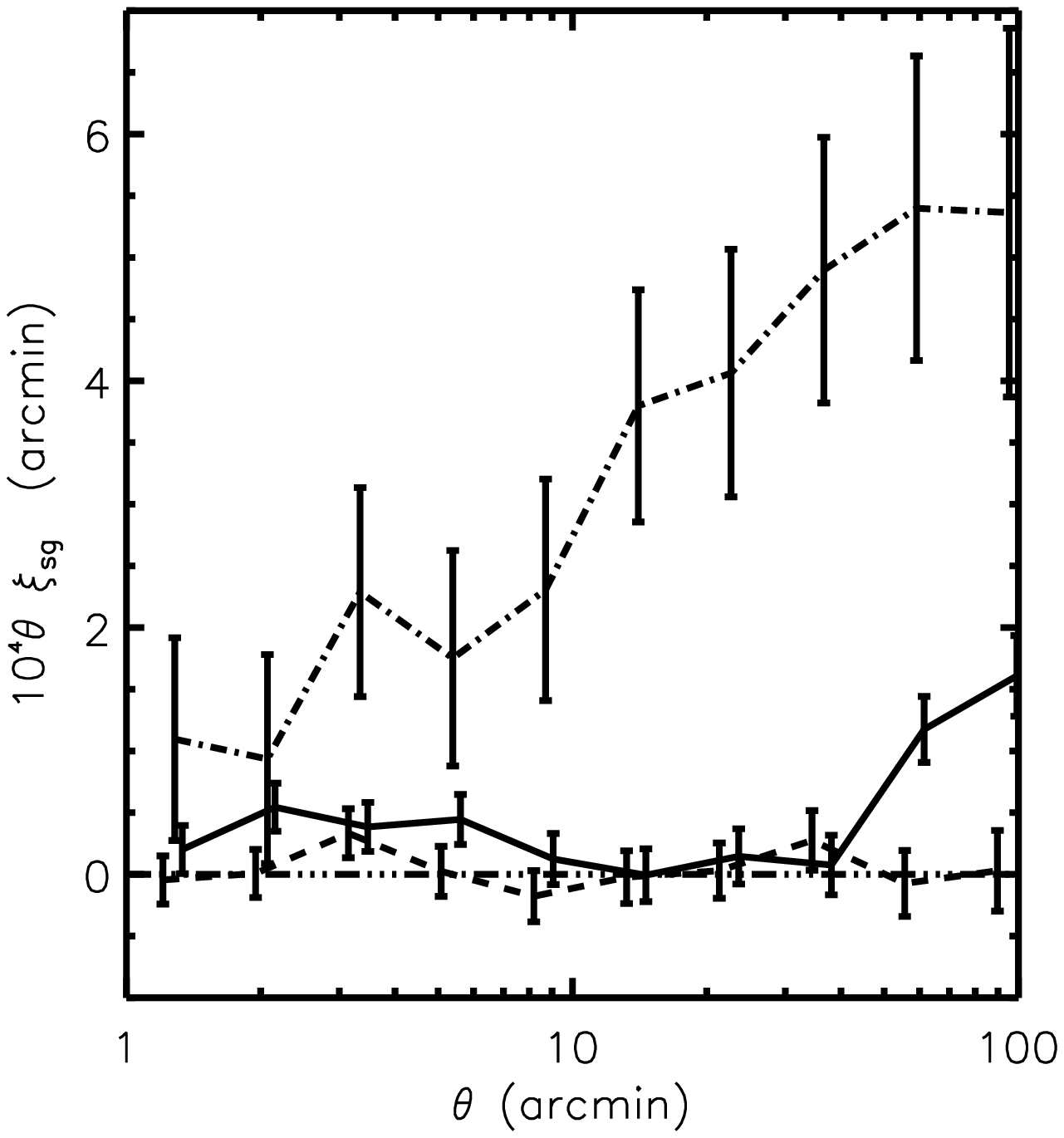} &
\includegraphics[width=3in,angle=0]{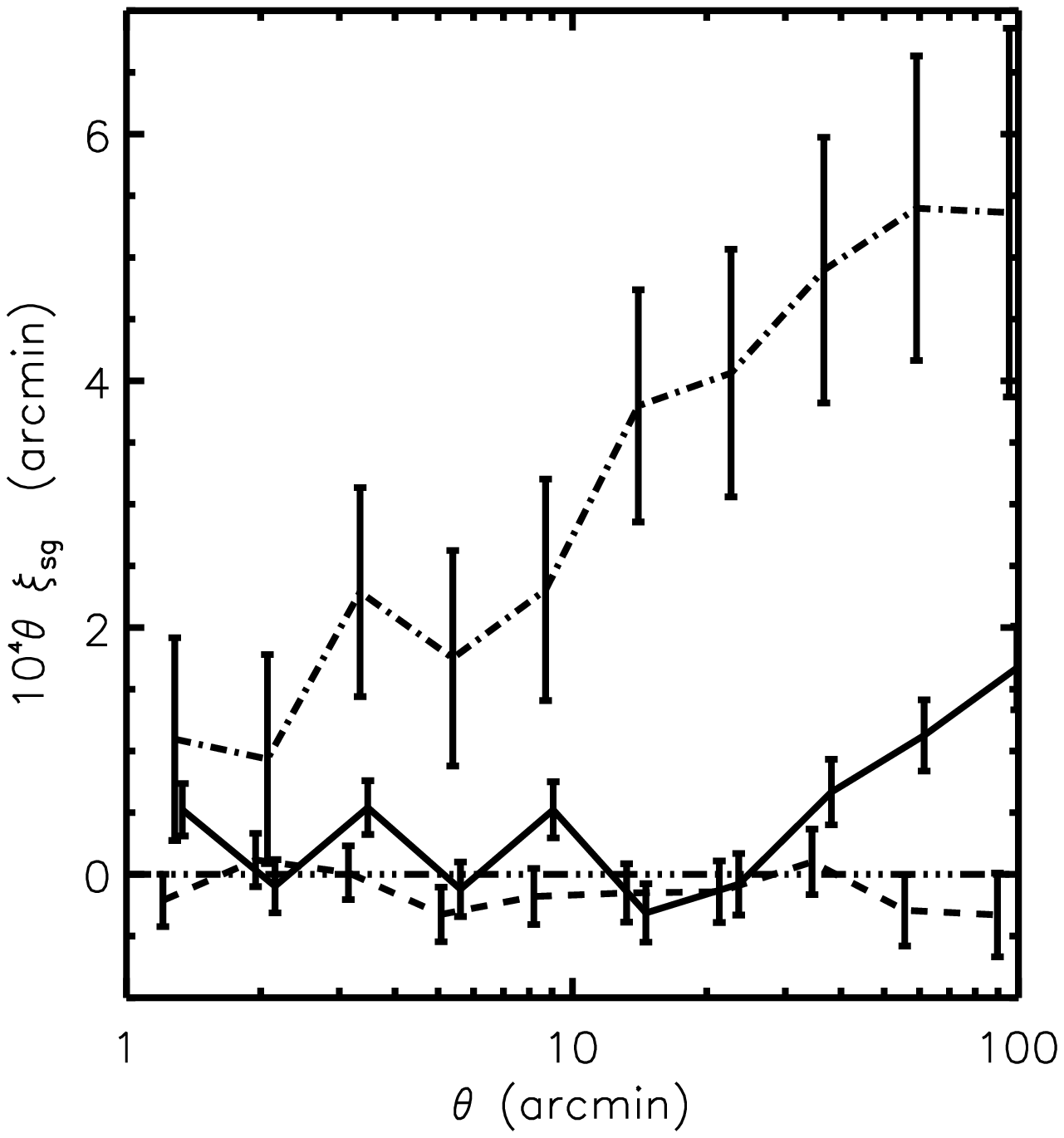} \\
\end{array}$
\caption{\label{fig:sg_ukidss}The cross-correlation of UKIDSS-selected star
  shapes with galaxy shapes, for the following pairs of bands:
  $(i,i)$ in the upper left, $(r,r)$ in the upper right, $(r,i)$ in
  the bottom left, and $(i,r)$ in the bottom right panels. All results
  are shown as $10^{4}\theta\xi$.  The
  $\langle e_1\,e_1\rangle$ correlation is the solid line, while the
  $\langle e_2\,e_2\rangle$ correlation is the dashed line. The
  dot-dashed line shows the expected cosmic shear
  $\langle e_+\,e_+\rangle$ shape-shape correlation for a survey of this
  depth and size, with shot-noise errors.  The triple dot-dashed lines
  shows the ideal value of zero for the star-galaxy correlations.}
\end{center}
\end{figure*}

As a test for this, we compute the star-galaxy shape correlation using
only those objects identified as stars in the manner described
above.  The results are shown in Fig.~\ref{fig:sg_ukidss}. As shown, for this population, the amplitude of the star-galaxy
correlation is significantly reduced below the star-galaxy
correlations. This is suggestive that some of the star-galaxy signal may arise from galaxy contamination of the star sample.
However, because the UKIDSS data does not cover the entire footprint of Stripe 82, this test is not conclusive.

After all of the above cuts have been applied, the final shape catalogue
consists of 1\,067\,031 {\it r}\,-band and 1\,251\,285 {\it i}\,-band
shape measurements, over an effective area of 140 and 168 square
degrees, respectively.

\subsubsection{Contamination of galaxy sample by stars}

The other type of contamination, that of the galaxy sample by stars,
will tend to dilute lensing statistics measured using our catalogue.
Because we wish to understand the contamination in a representative
sample of our galaxies (not just the ones bright enough to have a
match in the UKIDSS catalogue), we use a different strategy to
estimate this type of contamination.

The targeting photometry used for the DEEP2 survey comes from the
Canada-France-Hawaii Telescope (CFHT), and in the two DEEP2 fields on
Stripe 82, the typical seeing is 0.7--0.8 arcsec, nearly a factor of
two better than in our coadds \citep{2004ApJ...617..765C}.  The
catalogues from this imaging were publicly released\footnote{\tt
  http://deep.berkeley.edu/DR1/} as part of DEEP2 Data Release 1.  We
use the star versus galaxy classification for galaxies in the coadds
that have detections in the DEEP2 targeting photometry as a way to
estimate the stellar contamination in our catalogues in those fields.

We first match our galaxy catalogue against the DEEP2 targeting
photometry, finding matches for 96 per cent of our galaxies.  We then
eliminate those that are marked as bad data or saturated in the DEEP2
catalogues.  For the remaining objects, the star/galaxy separation
works as follows (see \citealt{2004ApJ...617..765C} for more details):
clearly extended objects are flagged as such, and we consider those as
secure galaxy detections.  Compact objects are assigned a quantity
`pgal' in the range $[0,1]$, representing the probability that the
object is a galaxy based on its colour and magnitude.  To estimate the
total stellar contamination, we sum the values of $(1-\mathrm{pgal})$
for all of our galaxies that matched against compact objects, and
compare that to the total number of compact and extended matches.  The
result is a stellar contamination of 1.7 per cent for both the $r$ and
the $i$ band lensing catalogues.

One issue in ``calibrating'' the stellar contamination analysis is
that the stellar density varies along the stripe. The Galactic
latitudes of the two DEEP2 fields are -54$^\circ$ and -56$^\circ$, but
the ``start'' of our stripe (RA 310$^\circ$, Dec 0$^\circ$) is at
$l=46^\circ$, $b=-24^\circ$. These low-RA fields are not only at lower
Galactic latitude, but also look inward toward the bulge. As a simple
test for this, we compared the ratio of the $19.5<i<21.5$ star density
to the source galaxy density in the DEEP2 fields (0.16) versus the
stripe as a whole (0.46). The ratio of these suggests that the DEEP2
field star abundances should be scaled up by a factor of 2.8 to be
representative of Stripe 82 as a whole.\footnote{About half of these
  stars are in the $310^\circ<$RA$<320^\circ$ range.} A potential
issue in this method of rescaling is that the stars of different
magnitude need not be distributed in the same way. To test for this,
we repeated the above computation for stars at fainter magnitudes,
with $22 < r < 22.5$ and found a rescaling factor of 1.4. This
suggests that the larger factor is conservative, but it is also
possible that the stellar density is being homogenized by galaxy
contamination of the stars. The true recalibration factor for DEEP2 is
probably greater than unity, but not larger than 2.8.

The statistical error on this contamination is $\sim 10$ per cent
(Poisson error); the systematic uncertainty in how it applies to a
real lensing analysis, given the strong variation in stellar density
across the stripe, is far larger. We therefore address the issue of
real corrections for a lensing analysis in Paper II.

\section{Discussion}\label{sec:discussion}

We have constructed deep, lensing-optimised, coadded imaging of the SDSS
equatorial stripe. The procedure is designed to enable the
construction of a catalogue suitable for weak lensing measurements by
suppressing the effects of PSF anisotropy on the measured galaxy
shapes below the level of statistical error achievable with a cosmic
shear survey on this Stripe.

The galaxy density of $\sim 2$ per arcmin$^2$ is relatively low for a cosmic shear survey.
However, it makes sense given our depth limits and large PSF of the SDSS,
even by ground-based standards. As a simple point of comparison, the CMC
\citep{2009A&A...504..359J} is commonly used to forecast galaxy yields for dark energy investigations.
The effective radius of the coadded PSF is 0.67 arcsec; for Gaussians one would then expect that our
cut on $R_2>0.333$ should correspond to a cut on effective radius of $r_{\rm eff}>0.67\sqrt{0.333/(1-0.333)}=0.47$ arcsec.
Using the 2011 August 15 update of the CMC,
and imposing this cut as well as $r<23.5$ and $i<22.5$ (observer frame at $A_r=0.141$), we forecast a galaxy density of $n=2.7$ arcmin$^{-2}$ and a mean source redshift $\langle z\rangle = 0.51$,
before any small-scale masking due to e.g. bright stars and bad columns. Therefore the final galaxy
yield is broadly consistent with the tools being used to design next-generation surveys.

\begin{figure}
\includegraphics[width=\columnwidth]{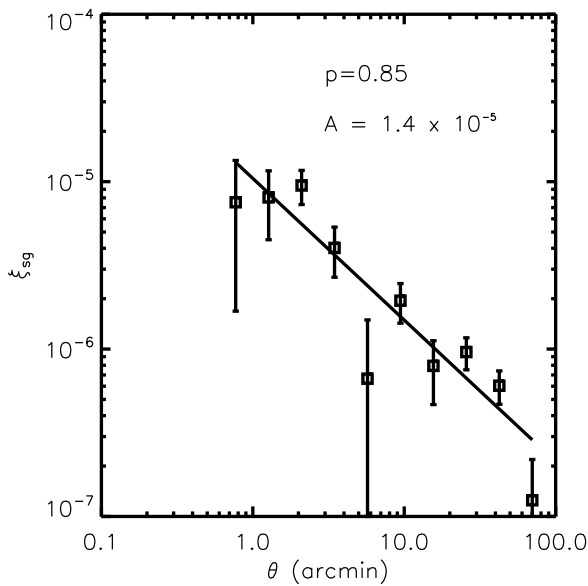}
\caption{Average of the $ri$, $ir$, $rr$, and $ii$ star-galaxy
cross-correlation functions, and the best-fit power-law from Eq.~(\ref{eq:sgpowlaw}).}
\label{fig:sg_avg}
\end{figure}

This procedure is successful if and only if it renders the PSF shape
distortions sufficiently small that they are negligible compared to
the statistical errors expected for a cosmic shear signal in this
survey. To estimate the amplitude and scale-dependence of the residual
PSF systematics, we have measured the star-galaxy and star-star ellipticity correlation
functions in our catalogue.
We now fit a power law of the form:
\begin{equation} \label{eq:sgpowlaw}
\xi_{sg}=A\theta^{-p}
\end{equation}
to the average of the four measured star-galaxy
cross-correlations, using the Poisson errors output by the correlation
function code. The best-fitting power law and average star-galaxy
correlations are shown in Fig.~\ref{fig:sg_avg}, with
$(A,p)=(1.4\times 10^{-5},0.85)$.

We compare the ratio of this best-fitting power law to the shot-noise errors
expected for a galaxy shape auto-correlation function for this survey. To
estimate the shot noise, we follow \citet{2002A&A...396....1S} to
calculate the statistical errors expected due to shot noise for a
$168$ square degree lensing survey with an effective source surface
density of 2 galaxies per arcmin$^2$:
\begin{eqnarray} 
{\rm Var}\left(\xi\right) &=&  \left(3.979 \times 10^{-9}\right)\left(\frac{\sigma_e}{0.3}\right)^4\left(\frac{\mathrm{Area}}{1\,{\rm deg}^2}\right)^{-1}
\nonumber\\ && \times  \left(\frac{n_\mathrm{eff}}{30\,{\rm arcmin}^{-2}}\right)^{-2} \left(\frac{\theta}{1\,{\rm arcmin}}\right)^{-2}.
\end{eqnarray}

\begin{figure}
\includegraphics[width=\columnwidth]{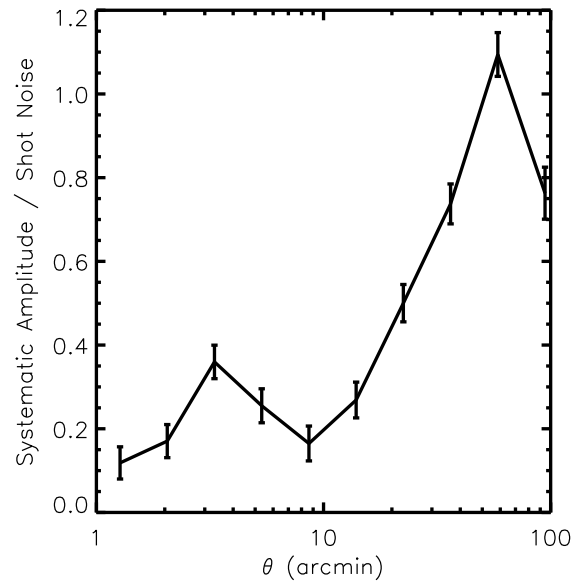}
\caption{Ratio of the best-fit star-galaxy cross-correlation power law
to the expected shot-noise errors for a cosmic shear measurement using
the catalogues described here. As the star-galaxy amplitude is only
poorly constrained, this should be taken as a rough indication of the
level of significance of the systematics.}
\label{fig:sys_ratio}
\end{figure}

The ratio of the systematics amplitude to the shot noise is shown as a
function of scale in Fig.~\ref{fig:sys_ratio}. From this, we can see
that PSF systematics for these data should be, on average, $50$ per
cent of the size of the statistical error budget for a cosmic shear
measurement with this catalogue; on degree scales, this becomes
comparable to the shot-noise errors.
As discussed above, this is an upper limit for three reasons: (1) imperfect star-galaxy
separation at the level of several to ten per cent can produce a star-galaxy
correlation signal in the absence of uncorrected PSF effect; (2) the
response of a galaxy shape to a PSF anisotropy is typically less than
unity; and (3) the Poisson error estimate will underestimate the true
variance of the signal on larger scales, where cosmic variance becomes
important.

In addition, masks defined as sets of pixels can introduce a shape
selection bias. We tested the effects of masking on the spurious shear
statistics during the catalogue-making step by applying a strict cut to
eliminate those regions of the coadd imaging with fewer than seven
contributing single-epoch images. Introducing this cut actually
{\em increased} the spurious shear amplitude; the star-galaxy correlations
in the presence of this more aggressive masking step reach an
amplitude of $10^{-5}$ at degree scales.

\begin{figure*}
\begin{center}
$\begin{array}{c@{\hspace{0.2in}}c}
\includegraphics[width=3in,angle=0]{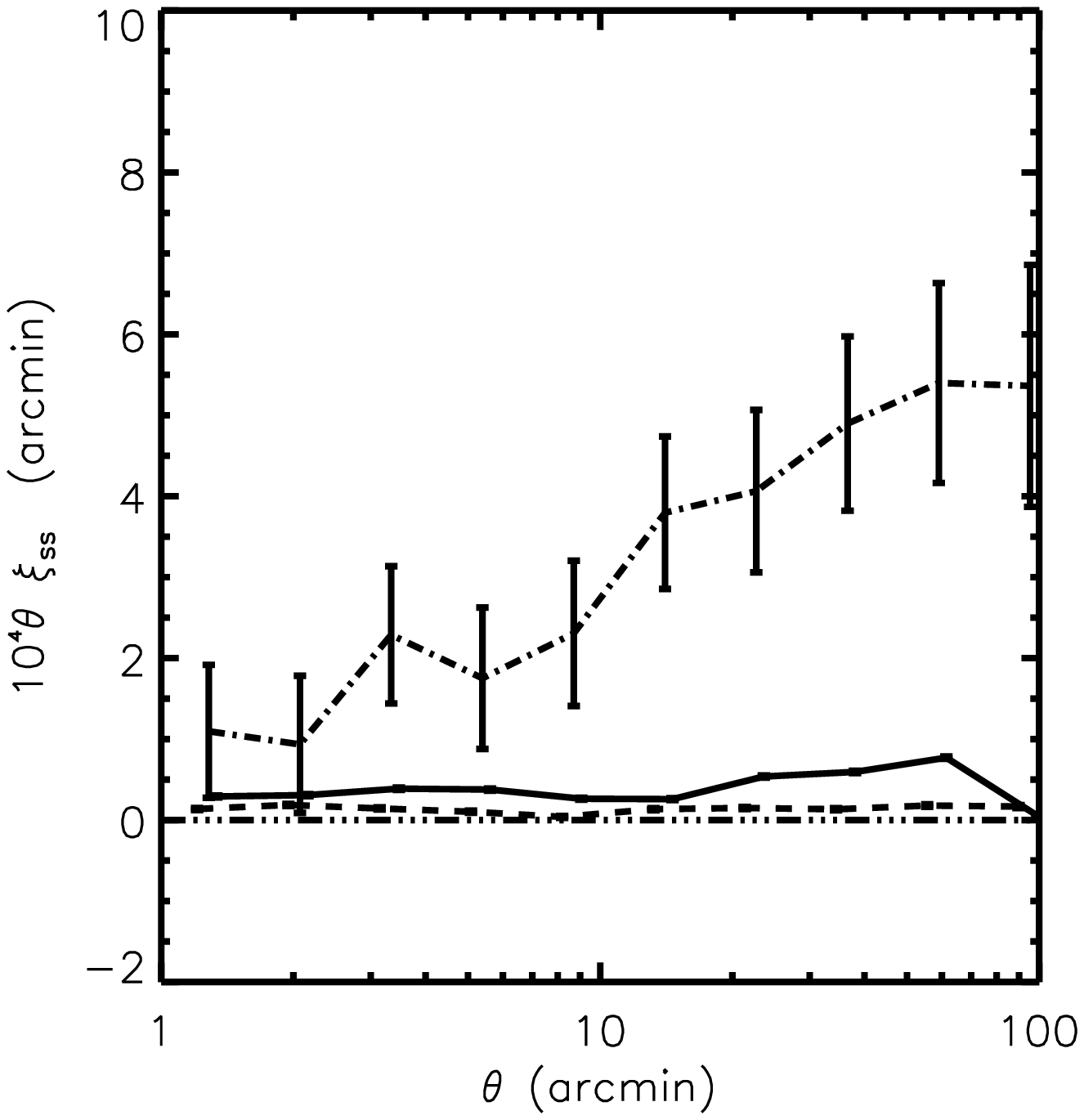} &
\includegraphics[width=3in,angle=0]{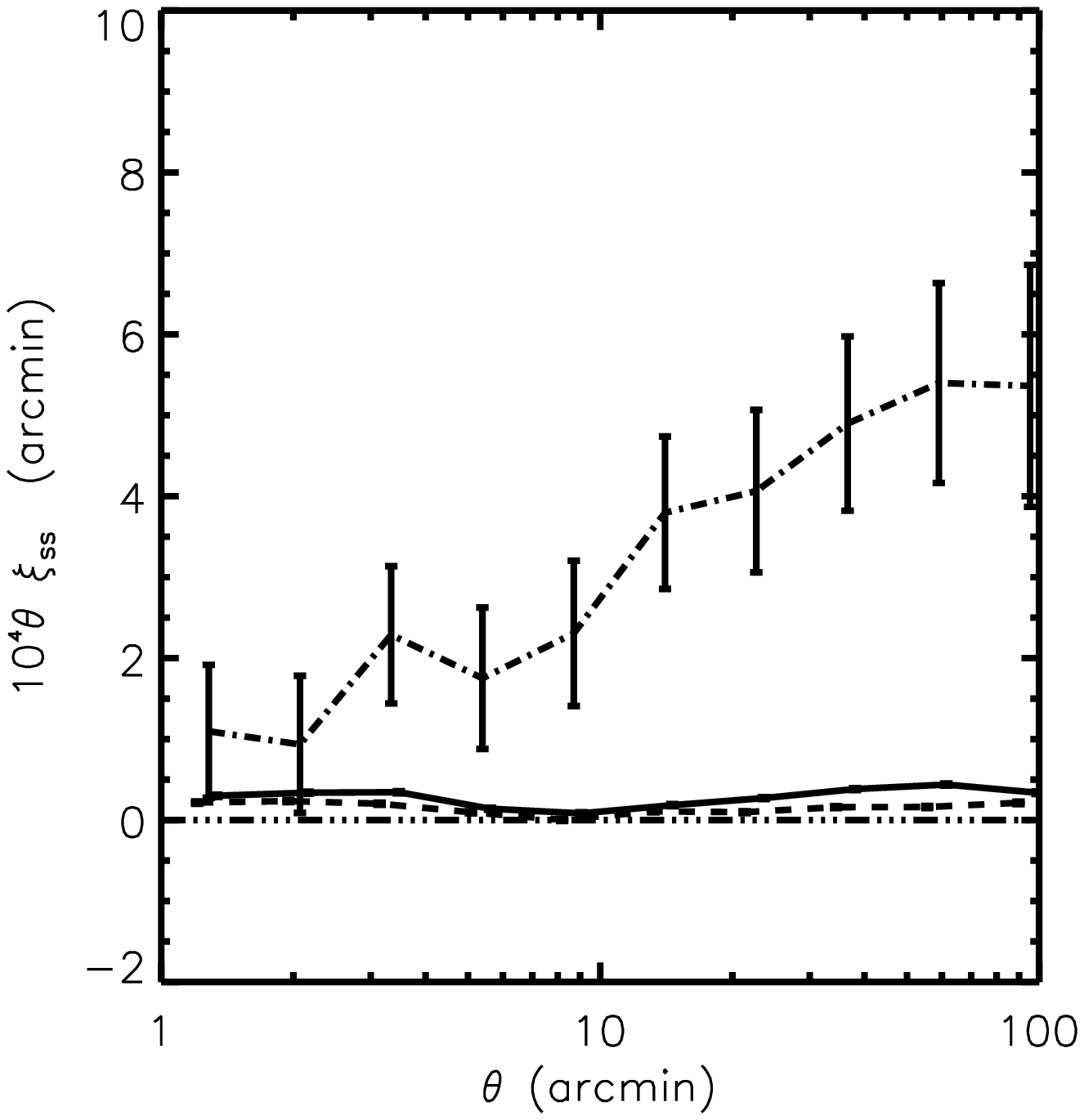} \\
\includegraphics[width=3in,angle=0]{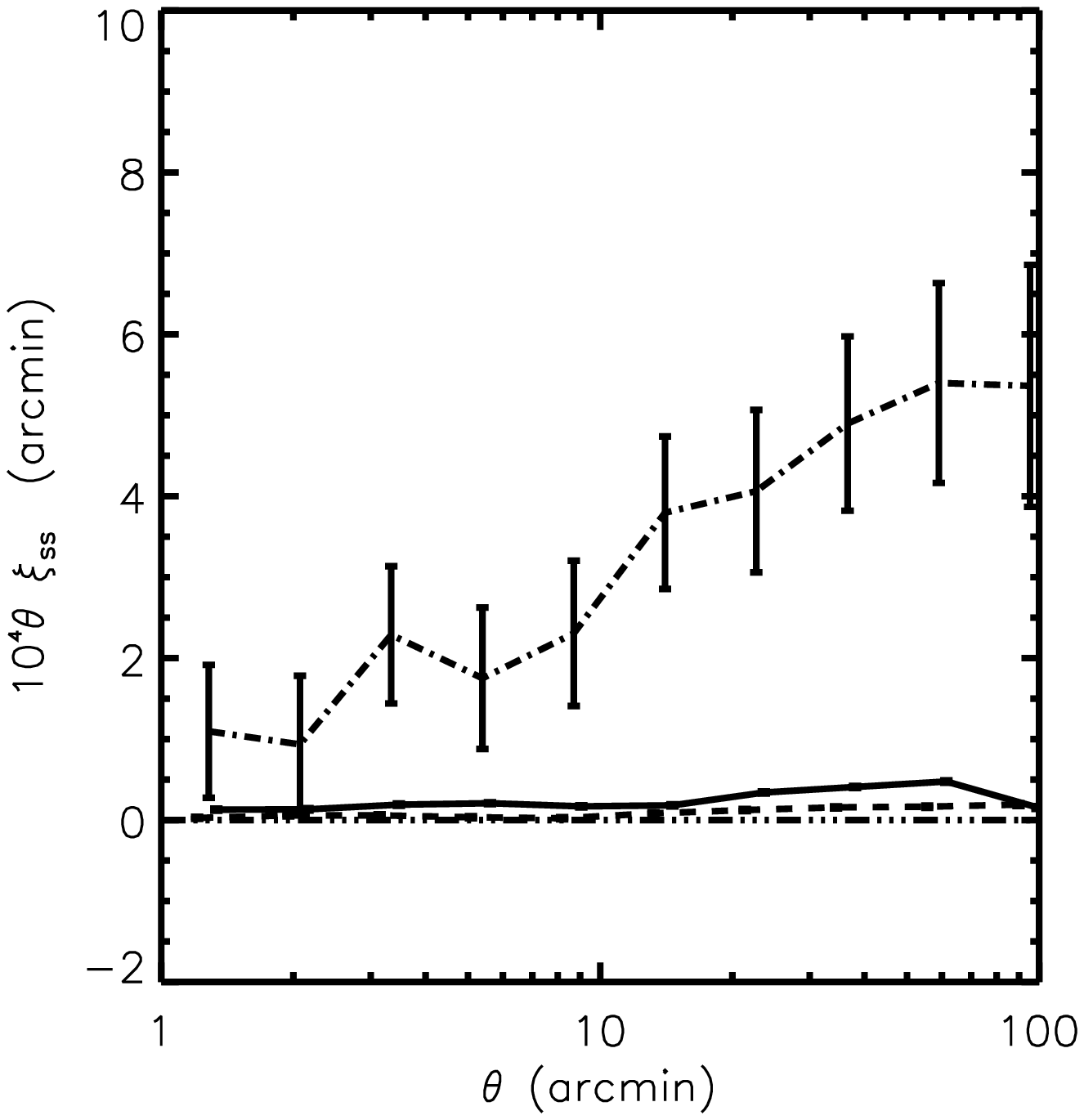} &
\includegraphics[width=3in,angle=0]{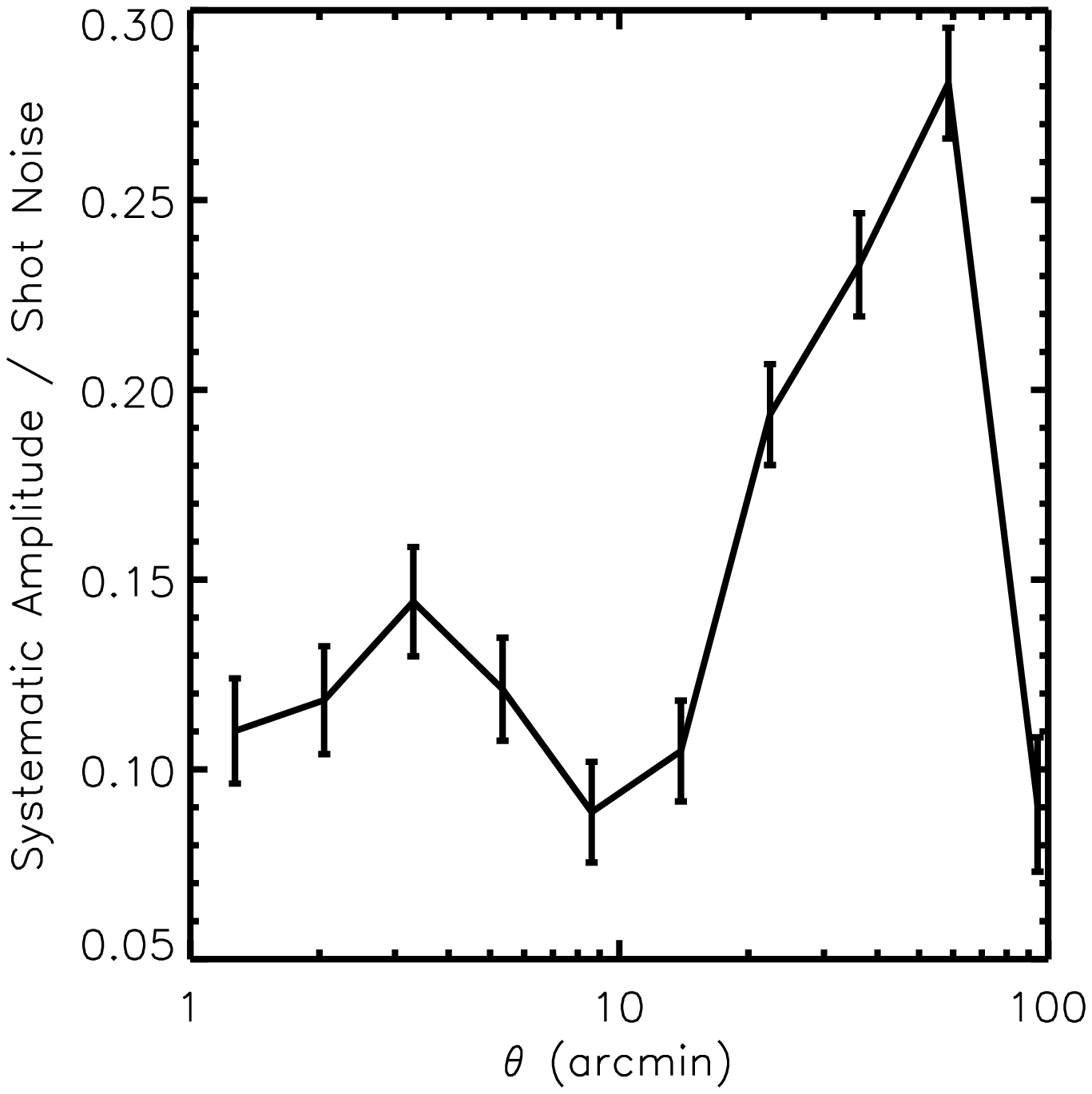} \\
\end{array}$
\caption{\label{fig:ss-all} The correlation functions of star shapes
  in the following pairs of bands: $(r,r)$ in top left, $(i,i)$ in top
right, and $(r,i)$ in bottom left panel.  All results
  are shown as $10^{4}\theta\xi$.  The
  $\langle e_1\,e_1\rangle$ correlation is the solid line, while the
  $\langle e_2\,e_2\rangle$ correlation is the dashed line. The
  dot-dashed line shows the expected cosmic shear
  $\langle e_+\,e_+\rangle$ shape-shape correlation for a survey of this
  depth and size, with shot-noise errors.  The lower right panel shows
  the mean stellar cross-correlation signal as a fraction of the
  expected Poisson error for a cosmic shear measurement using this
  catalogue  The triple dot-dashed lines
  shows the ideal value of zero for the star auto-correlations.
}
\end{center}
\end{figure*}

The relative contributions of mask selection and PSF anisotropy biases
can be ascertained from the relative amplitudes of the star-star and
star-galaxy correlation functions. A PSF anisotropy will produce a
similar signal in both metrics. The stellar shape dispersion and
typical stellar size is much
smaller than that of the galaxies, so a selection bias will produce a
much larger systematics signal in the star-galaxy correlation function
than in the star-star correlation functions. This is indeed the case,
as shown in Fig.~\ref{fig:ss-all} --
substantial evidence that mask selection bias will be a significant
fraction of the systematic error budget. Excluding objects near the
boundaries of masked regions on the basis of their centroid positions
could remove this effect; however, as Fig.~\ref{fig:sys_ratio}
shows, the statistical errors should dominate for this catalogue, so
reducing the catalogue further at this stage would not improve the
quality of a final cosmic shear measurement. We also considered trying
to simulate and subtract the masking bias. Ultimately, however, we settled
on a more empirical approach: as described in detail in Paper II, the
galaxy shear autocorrelation function used in the final analysis determines
and subtracts the mean $e_1$ in each declination bin.

These results suggest that a cosmic shear analysis that is
statistics-limited is possible with these data. We have shown that the
effects of the point-spread function are small compared to the
statistical errors. The mask selection bias is larger, but still on
average significantly smaller than the expected statistical errors. A
full analysis involving the source redshift distribution, shear
calibration, and the cosmological implications of the two-point
statistics of these data will follow in Paper II.

The systematics floor for the rounding kernel method we have employed
here is set by the SDSS PSF model. Inaccuracies in this PSF model are
documented both here (Fig.~\ref{fig:binnedstars}) and in other work
\citep{2011arXiv1110.4107R}. Coherent variations in the PSF model
errors in both components across the camera columns are visible with a
characteristic amplitude of $2\times 10^{-3}$. Aside from the very
striking and atypical effect seen in the $r$ band in camcol 2, it is
likely that the shortcomings of the polynomial interpolation method
employed in {\sc Photo} play an important role here, as documented in
\cite{2011arXiv1110.2517B} for more general simulated ground-based
data. As this is close to the level of residual PSF systematics seen
in our final lensing catalogue, it is very likely that an improvement in
the underlying model construction would allow the rounding kernel
method deployed here to achieve a greater level of systematics
control.

The masking problem that we have identified is not extensively treated in the literature; to
the knowledge of the authors, it has not been taken into account in
existing studies. It is standard in modern photometric pipelines to
define the survey mask and object rejection algorithms in terms of
sets of pixels, rather than (for example) galaxy centroids, which is
the ultimate source of the masking bias we see here. This effect will
be important to take into account in the photometric pipeline
construction in the next generation of lensing measurements. If possible,
masking-related biases (and more generally, survey uniformity) should also be addressed at the observing strategy
level. In this regard, the SDSS Stripe 82 technique of scanning the sky along the
same guiding great circle many times, while appropriate for supernovae or
transient searches, was highly non-optimal from the perspective of producing a
uniform quality co-added image, since bad columns and other defects
always occur at the same positions. Even dithering successive runs in the cross-scan (declination) direction by of order 10 arcsec would have
helped this project enormously.

In Paper II, we will use the catalogue described here to measure
cosmic shear.  While this work was underway, we learned of a parallel
effort by Lin et al. (2011). These two efforts use different methods
of coaddition and different sets of cuts for the input images and
galaxies; what they have in common is their use of SDSS data (not
necessarily the same set of runs) and their use of the SDSS {\sc
  Photo} pipeline for the initial reduction of the single epoch data
and the final reduction of the coadded data (however, they use
different versions of {\sc Photo}).  Using these different methods,
both groups have attempted to extract the cosmic shear signal and its
cosmological interpretations. We have coordinated submission with them
but have not consulted their results prior to this, so these two
analysis efforts are completely independent, representing an extreme
version of two independent pipelines.

The PSF correction method described here is suitable for deployment in
the next generation of weak lensing surveys, or generally any survey with many fully sampled images of the same region.\footnote{See \citet{2011ApJ...741...46R} for an algorithm similar in spirit to that used here, but that could be applied to undersampled data.} All of these surveys will
include multi-epoch data over their full footprint. In many cases the image
quality (as measured by PSF isotropy and size) distribution may be fairly
narrow, since the ``shape measurement'' bands will be acquired in the best seeing conditions.
In these cases, the rounding kernel method will result in little loss
of information.  


\section*{Acknowledgments}

We thank Gary Bernstein, Alison Coil, Tim Eifler, Jim Gunn, Mike
Jarvis, Alexie Leauthaud, Reiko Nakajima, Jeff Newman, Nikhil
Padmanabhan, and Barney Rowe for many useful discussions about this
project.  We thank Kevin Bundy for allowing us to use preliminary
versions of his UKIDSS-SDSS colour-matched catalogue.

E.M.H. is supported by the US Department of Energy's Office of High Energy Physics (DE-AC02-05CH11231).
During the period of work on this paper, C.H. was supported by the US
Department of Energy's Office of High Energy Physics (DE-FG03-02-ER40701 and DE-SC0006624), the US National Science
Foundation (AST-0807337), the Alfred P. Sloan Foundation, and the
David \& Lucile Packard Foundation. R.M. was supported for part of
the duration of this project by NASA
through Hubble Fellowship grant \#HST-HF-01199.02-A awarded by the
Space Telescope Science Institute, which is operated by the
Association of Universities for Research in Astronomy, Inc., for NASA,
under contract NAS 5-26555.  U.S. is supported by the DOE, the Swiss National Foundation under contract 200021-116696/1 and WCU grant R32-10130.

Funding for the SDSS and SDSS-II has been provided by the Alfred P. Sloan Foundation, the Participating Institutions, the National Science Foundation, the U.S. Department of Energy, the National Aeronautics and Space Administration, the Japanese Monbukagakusho, the Max Planck Society, and the Higher Education Funding Council for England. The SDSS Web Site is {\tt http://www.sdss.org/}.

The SDSS is managed by the Astrophysical Research Consortium for the Participating Institutions. The Participating Institutions are the American Museum of Natural History, Astrophysical Institute Potsdam, University of Basel, University of Cambridge, Case Western Reserve University, University of Chicago, Drexel University, Fermilab, the Institute for Advanced Study, the Japan Participation Group, Johns Hopkins University, the Joint Institute for Nuclear Astrophysics, the Kavli Institute for Particle Astrophysics and Cosmology, the Korean Scientist Group, the Chinese Academy of Sciences (LAMOST), Los Alamos National Laboratory, the Max-Planck-Institute for Astronomy (MPIA), the Max-Planck-Institute for Astrophysics (MPA), New Mexico State University, Ohio State University, University of Pittsburgh, University of Portsmouth, Princeton University, the United States Naval Observatory, and the University of Washington.

\end{document}